\documentclass[10pt]{article}
\usepackage{appendix}
\usepackage{amsmath}
\usepackage{amssymb}
\usepackage{graphicx}
\usepackage{cite}
\usepackage{color}

\usepackage{xcolor}

\usepackage[labelfont=bf,labelsep=period,justification=raggedright]{caption}

\usepackage{rotating}

\makeatletter
\renewcommand{\@biblabel}[1]{\quad#1.}
\makeatother

\date{}

\begin{document}

\begin{flushleft}
{\Large \textbf{The evolutionary dynamics of protein-protein
interaction networks inferred from the reconstruction of ancient
networks} }
\\
Yuliang Jin$^{1,\dag}$, Dmitrij Turaev$^{2,\dag}$, Thomas Weinmaier$^{2,\dag}$, Thomas Rattei$^{2,\ddag}$, Hern\'{a}n A. Makse$^{1,\ast}$
\\
{\bf 1} Levich Institute and Physics Department, City College of New York, New York, New York, United States of America
\\
{\bf 2} Department of Computational Systems Biology, University of Vienna, Vienna, Austria
\\
$\dag$ These authors contributed equally.\\
$\ast$ E-mail: hmakse@lev.ccny.cuny.edu\\
$\ddag$ E-mail: thomas.rattei@univie.ac.at\\
\end{flushleft}

\section*{Abstract}
Cellular functions are based on the complex interplay of proteins, therefore the structure and dynamics of these protein-protein interaction (PPI) networks are the key to the
functional understanding of cells. In the last years, large-scale PPI networks of several model organisms were investigated. A number of theoretical models have been developed to
explain both, the network formation and the current structure. Favored are models based on duplication and divergence of genes, as they most closely represent the biological foundation of
network evolution. However, studies are often based on simulated instead of empirical data or they cover only single organisms. Methodological improvements now allow the analysis of PPI
networks of multiple organisms simultaneously as well as the direct modeling of ancestral networks. This provides the opportunity to challenge existing assumptions on network
evolution. We utilized present-day PPI networks from integrated datasets of seven model organisms and developed a theoretical and bioinformatic framework for studying the
evolutionary dynamics of PPI networks. A novel filtering approach using percolation analysis was developed to remove low confidence interactions based on topological constraints. We
then reconstructed the ancient PPI networks of different ancestors, for which the ancestral proteomes, as well as the ancestral interactions, were inferred. Ancestral proteins were
reconstructed using orthologous groups on different evolutionary levels. A stochastic approach, using the duplication-divergence model, was developed for estimating the probabilities of
ancient interactions from today's PPI networks. The growth rates for nodes, edges, sizes and modularities of the networks indicate multiplicative growth and are consistent with the
results from independent static analysis.
Our results support the duplication-divergence model of evolution and indicate fractality and multiplicative growth as general properties of the PPI network structure and
dynamics.


\section*{Introduction}

A living cell relies on a wide network of protein-protein
interactions (PPIs) of structural and functional relevance,
therefore the understanding of cell function is intrinsically tied
to the understanding of this network. Technical advances in
molecular and cellular biology and bioinformatics enabled
extensive studies on protein-protein interaction networks (PIN)
during the last decade. While a significant amount of data was
collected during this time, theoretical analyses were focused on
PINs from very few model organisms. Little is known about the
comparability of results from different organisms as well as their
transferability \cite{mika_protein-protein_2006,
zinman_biological_2011}. General theoretical models explaining the
formation, function and emerging properties of biological networks
however often lack the connection to empirical data, making it
difficult to validate the models \cite{gibson_improving_2011}.
Here we improve network theory for studying the evolutionary
dynamics of PIN in multiple organisms.\\

{\bf Experimental determination of protein-protein interaction networks}\\
Multiple experimental methods for measuring PPI networks have been
developed, like the yeast two-hybrid screen (Y2H)
\cite{Fields2005, suter_two-hybrid_2008, koegl_improving_2007},
the tandem affinity purification/mass spectrometry (TAP-MS)
\cite{Gavin2006, Krogan2006, Wodak2009} and the protein-fragment
complementation assay \cite{tarassov_vivo_2008}. Each method has
specific characteristics and limitations and therefore can provide
only an incomplete view of the biological reality. For example,
while TAP-MS detects stable complexes, weak and transient
interactions are more readily detected by Y2H \cite{Yu2008}.
The precise determination of the error rates is difficult. For example, for Y2H experiments, estimates
range from 10\% to over 50\% for the false positive rate and from
30\% to 90\% for the false negative rate
\cite{von_mering_comparative_2002,huang_where_2007}. Furthermore,
a bias is introduced by variations in the details of the Y2H
protocol, such as the vectors used and the nature of the
re-constituted transcription factor
\cite{braun_experimentally_2009, rajagopala_benchmarking_2009}.
For these reasons, the overlap between different studies is often
small \cite{von_mering_comparative_2002, Yu2008,
koegl_improving_2007}. Possible approaches that can be applied for
the selection of reliable interactions are reproducability,
promiscuity, indirect support, conservation and topology
\cite{collins_toward_2007, koegl_improving_2007}, whereas the best
suited approach depends on the specific dataset.\\

Due to the volume of work and the methodological difficulties, genome-wide interactome studies were so far performed for only a limited number of organisms, among others \textit{S. cerevisiae} \cite{Yu2008}, \textit{H. sapiens} \cite{rual_towards_2005} and \textit{A. thaliana} \cite{ath_consortium_evidence_2011}. The results of these large-scale experiments and many other studies are collected in a number of databases like Mint \cite{licata_mint_2012}, DIP \cite{salwinski_database_2004}, BioGrid \cite{breitkreutz_biogrid_2008} and IntAct \cite{kerrien_intact_2012}.  These resources are partially redundant and use different database schemes, scores and identifiers. Integrating data from these sources for comprehensive analysis is therefore non-trivial. This problem is tackled e.g. by the STRING database, which incorporates different evidence sources for both physical and functional PPIs \cite{szklarczyk_string_2011}.\\

{\bf Structure and topology of protein-protein interaction networks}\\
For the characterization of the network structure, measures from network theory, like
node degree, clustering coefficient or shortest path are
used \cite{Barabasi2004}. Based on these measures, observed
networks can be assigned to different topological categories like
random\cite{Bollobas1985}, small-world\cite{Watts1998}, hierarchical\cite{Ravasz2002},
fractal\cite{Song2005}, and scale-free \cite{albert_error_2000, Barabasi2004}. \\

PPI networks often show the small-world property, namely a
short path length between any two nodes. The additional shortcuts
in small-world networks affect the modularity, as well as the path
length between proteins, and might for example influence signal
transduction \cite{Watts1998}. For small-world networks the
scaling of the number of nodes and the average distance is
exponential. It has also been shown that many complex networks
show a scale-free topology, with the degree distribution following
a power-law with the degree exponent $\gamma$ \cite{Albert2002,
Barabasi1999}. A scale-free topology results in a high robustness
of the network against perturbations \cite{albert_error_2000}.\\

PPI networks have also been shown to exhibit a highly modular structure, that is they contain substructures which are highly interconnected but have only few connections to nodes outside the module\cite{Wagner2001, Barabasi2004}. The modular organization represents the higher-order correlations of the network structure beyond average properties, and has attracted great attention because it is closely related to the network functionality and robustness. For example, it has been shown that the modularity increases the overall robustness of the network by limiting the effect of local perturbations \cite{Barabasi2004, Song2006, Gallos2007}. Along with the modular organization, the fractal and
self-similar feature is empirically observed in many biological
networks, such as the protein PPI networks\cite{Song2005}, the
biochemical reactions in metabolism \cite{Song2005}, and the human
cell differentiation networks \cite{Galvao2010}. The fractal
network is characterized by a power-law scaling between the
average distance and the number of nodes, as well as an
organization of hubs which are preferentially connected to small
degree nodes (disassortativity) rather than other hubs
\cite{Song2006,Goh2006}.\\

{\bf Dynamics and evolution of protein-protein interaction networks}\\
The primary source of node evolution is assumed to be the duplication of single genes, groups of
genes or whole genomes followed by divergence of duplicated genes
\cite{presser_evolutionary_2008, Ohno1970, Li1997, Patthy1999,
Taylor2004}, whereas link evolution has been modeled by different
mechanisms such as random rewiring \cite{wagner_how_2003} and
preferential attachment \cite{Barabasi1999}. Network rewiring can for example
be studied by tracking the evolution of network motifs after a
whole-genome duplication event with subsequent divergence
\cite{presser_evolutionary_2008}. The change in protein-protein
interactions between related species was shown to be substantially
lower than the rate of protein sequence evolution
\cite{qian_measuring_2011-1}. These general considerations of
network evolution indicated that frequently observed topological
features like scale-free degree distribution (and preferential
node attachment) are explained by mechanisms of network growth
rather than by natural selection \cite{wagner_how_2003}. Later
studies demonstrated that the evolutionary conservation and the
topology of networks are readily explained by exponential
duplication/divergence dynamics (DDD) \cite{Evlampiev2007,
Evlampiev2008}. \\

Mathematical models based on these mechanisms \cite{Sole2002, Kim2002, Chung2002, Evlampiev2008, Vazquez2003} often well reproduce the observed degree distribution $P(k)$ from
numerical simulations of random graphs or analytical solutions of the asymptotic behaviors. However, two networks with the same $P(k)$ can have a totally different modular structure
which is determined by higher-order correlations, and not captured by the simple degree distribution $P(k)$. Furthermore, the simulated graphs generally do not correspond to the
history of real networks, and the comparisons with experimental data are usually ambiguous as the parameters used in the models are difficult to measure directly.\\ Later studies
utilize multiple approaches based on extant interaction networks for the explicit reconstruction of ancient networks which are then used to construct evolutionary arguments. Parsimony
methods are motivated by the idea that network evolution is best explained by the least evolutionary changes \cite{mirkin_algorithms_2003, patro_parsimonious_2012}, whereas probabilistic methods reconstruct
ancient networks of maximum likelihood \cite{Navlakha2011, zhang_refining_2010}. Integrating also phylogenetic information of the proteins represents their evolution more closely and therefore can further
improve the accuracy of the reconstructed networks \cite{pinney_reconstruction_2007, gibson_reverse_2009, Dutkowski2007}. One of the most recent methods allows parsimonious reconstruction of multiple evolutionary events and at the same time it makes fewer assumptions compared to previous studies\cite{patro_parsimonious_2012}.
Dutkowski et al \cite{Dutkowski2007} suggested to use clusters of orthologous groups (COGs) to reconstruct ancestral proteins and ancestral interactions. Here we prefer the concept of COGs for reconstructing ancestral PPI network nodes, as it has been shown to be very robust and applicable even to evolutionarily distant genomes. COGs are therefore well established in comparative genomics (reviewed in \cite{koonin_orthologs_2005}).\\

Most hitherto existing studies on network evolution were
conducted on PPI networks of single organisms - mostly yeast, due to the rich amount of data - or on PPI networks of a small number of organisms. Integration of further organisms into
evolutionary investigations allows for more general and more reliable statements on evolutionary principles. Facilitating the phylogenetic history of present-day proteins along with
orthologous relationships between proteins offers a powerful possibility for the reconstruction of ancient proteins \cite{Kunin2004}. However, no similar concept
exists for the inference of ancient interactions based on extant ones, therefore an underlying evolutionary model is necessary for their reconstruction.\\

The availability of large-scale PPI data for different species
renders it now possible to study the dynamics of PPI networks of
multiple species comprehensively by a novel approach combining advanced network
theory and bioinformatics. Relying on the rich body of previous
theoretical work as discussed above, we have established a
theoretical framework by which we explicitly reconstruct and
analyze ancestral PPI networks. The framework is based on clusters
of orthologous groups for the genome-wide representation of ancestral
proteomes on different taxonomic levels and a new stochastic model
describing the duplication-divergence processes. The assumption of
fractal topology of PPI networks, well justified by previous
research, allows to properly handle the noisy and erroneous input
data and to reduce the parameter space for the modeling of
ancestral PPIs. The analysis of the degree distribution $P(k)$
separates different species into two groups, characterized by a
power-law (scale-free) distribution (\textit{M. musculus},
\textit{C. elegans}, \textit{D. melanogaster} and \textit{E.
coli}), and an exponential distribution (\textit{S. cerevisiae},
\textit{H. sapiens} and \textit{A. thaliana}). Irrespective of
this, we find that their network topologies can be unified under
the framework of scaling theory and characterized by a set of
unique scaling exponents. The evolution of PPIs based on DDD can
be modeled using two parameters, describing the probability for
retaining an interaction after a duplication and the probability
of a \emph{de novo} creation of an interaction respectively. These
iterative duplication events due to DDD imply a multiplicative
growth of nodes, interactions and average path length that can be
described by dynamic growth rates. The growth rates
were obtained directly from the reconstructed networks. We observed that they
are in agreement with the mechanisms of multiplicative growth, which was
previously suggested in a theoretical study \cite{Song2006}. They
are also in good agreement with the static measurements of the
present-day
networks.\\

\section*{Results}

{\bf A uniform database allows for the comprehensive analysis of present-day interactomes}

\begin{figure}[!ht]

\begin{center}
\hbox{A \includegraphics[scale=0.5, angle=-90]{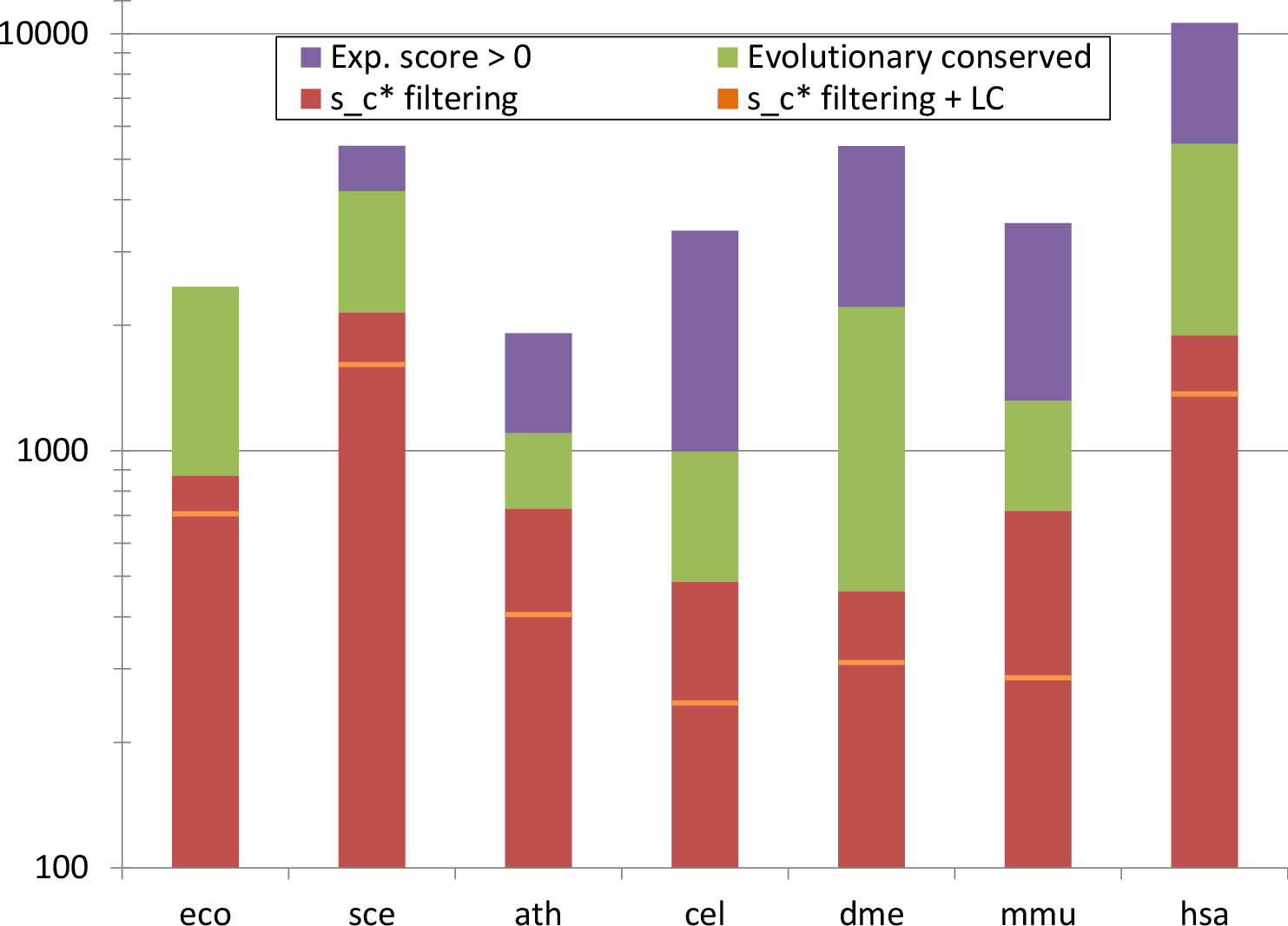}}
\medskip
\hbox{B \includegraphics[scale=0.5, angle=-90]{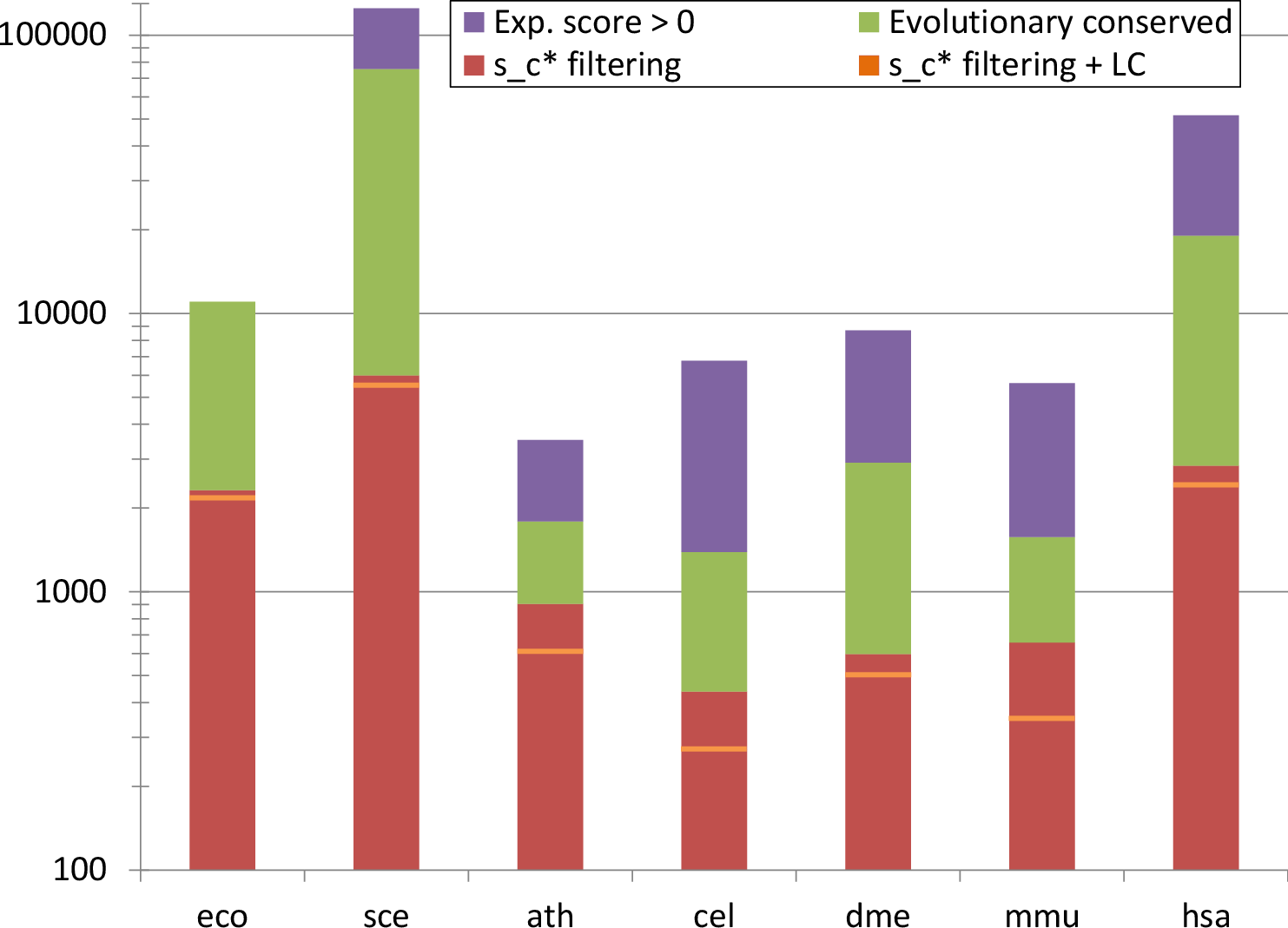}}
\end{center}
\caption{{\bf Input data overview.} The numbers of proteins (nodes) and interactions extracted from STRING at each filter step before construction of the protein-protein interaction
networks. Numbers are show on log-scale. (A) Number of nodes. (B) Number of interactions. Violet: STRING experimental score $> 0$, green: conserved on all evolutionary levels, red:
after filtering at $s_c^*$, orange bars: after filtering at $s_c^*$ considering only largest (connected) component (LC); the largest component is necessary for the topological
analysis} \label{fig:inputData}
\end{figure}

To elucidate the broad principles governing the structure and the
evolution of PPI networks, the most comprehensive and reliable
data for as many species as possible are necessary. This is why
the integrative database STRING \cite{szklarczyk_string_2011} was
chosen as the uniform source for physical protein-protein
interactions. Besides functional interactions, which are not
considered in this study, STRING provides physical PPIs for many
species. For this study we selected seven species having the
highest number of physical interactions in STRING and representing
different lineages in eukaryotes and bacteria
(Table~\ref{tab:organisms}). To construct high-quality physical
PPI networks from these data, a number of filtering steps was
performed. First, interactions without direct experimental
evidence for the respective organism were removed from the
analysis. This guaranteed that neither functional nor predicted
physical interactions (interologs) were included in network
construction. Second, proteins that are not contained in
orthologous groups on all evolutionary levels defined by the
eggNOG database \cite{Muller2010} for the respective organism were
excluded. This step removes all lineage specific proteins and
provides consistent sets of nodes for the subsequent modeling of
ancient PPI network (see below). Third, a threshold for confidence
scores was introduced to separate high-confidence from
low-confidence interactions, which were excluded from further
analysis. The confidence scores are very differently distributed
in the seven organisms of our study
(Figure~\ref{fig:boxplot_allOrganisms_expScore}). Application of a
uniform threshold score (e.g. 700) as generally suggested by
STRING \cite{szklarczyk_string_2011} would select very different
fractions of the interaction data. As all further results of this
study rely on the quality and unbiased selection of the
interactions from STRING, we evaluated the effect of different
score thresholds on the structure of the resulting networks. It is
known that PPI networks are invariant or self-similar under a
length-scale transformation \cite{Song2005}. This basic assumption
about the structure of the resulting networks was therefore
utilized to determine the optimal cutoff scores for each organism
by three independent methods (see Materials and Methods, and
Figure~\ref{fig:percolation}): percolation analysis, the Maximum
Excluded Mass Burning (MEMB)\cite{Song2007} and the
renormalization group approach \cite{Rezenfeld2010}. The
percolation analysis allowed to identify a point of percolation
transition, at which a giant connected component first appears.
This point of percolation transition was determined individually
for each organism. At the point of percolation transition, the
structure of the resulting networks changes from small-world to
self-similar. The box-covering algorithm MEMB and the
renormalization group approach served to validate the percolation
analysis by confirming the self-similar structure of the resulting
networks. Score thresholds between 400 (\textit{A. thaliana}) and
980 (\textit{S. cerevisiae}) were obtained for the different
organisms (Figure~\ref{fig:boxplot_allOrganisms_expScore} and
Table~\ref{tab:organisms}). The filtering always removed the
majority of proteins and interactions (Figure~\ref{fig:inputData}
and Table~\ref{tab:completeInputdataOverview}).\\

For the topological characterization of the seven PPI networks we
selected the largest connected component of every network.
The application of the MEMB algorithm revealed a power-law relationship between the minimum number of boxes $N_B$ and the box diameter $\ell_B$ (Equation~\ref{eq:fractal}), which is
typical for self-similar networks as shown in \cite{Song2007}. In this algorithm, $d_B$ is the fractal dimension which characterizes the self-similarity between different topological
scales of the network. It is known that the fractal dimension $d_B = 2$ for random Erd\"{o}s-R\'{e}nyi (ER) network at percolation \cite{Bunde1991}. Our results suggest that the PPI
networks have modular structures with correlated rather than random connections, since their values of $d_B$ (Table~\ref{tab:exponents_static}) are different from the one predicted by
the random percolation theory. Since the degree of modularity depends on the scale $\ell_B$, the modularity exponent $d_M$ was calculated which can be used to compare the strength of
modularity between dissimilar networks (Equation~\ref{eq:dm} and Figure \ref{fig:modularity}). The degree of modularity of the networks ranges from low ($d_M = 1.3(4)$) for \textit{E.
coli} and \textit{S. cerevisiae} to high for \textit{A. thaliana} ($d_M = 2.1(2)$), \textit{M. musculus} and \textit{H. sapiens} (both $d_M = 2.0(1)$) (Table
~\ref{tab:exponents_static}). Since the trivial case of a regular lattice in $d$ dimensions gives $d_M=1$, modularity exponents larger than one indicate a larger degree of modularity.
Besides the fractality, another important topological measure is the distribution of degrees $P(k)$. For many complex networks, $P(k)$ has a power law distribution with degree exponent
$\gamma$ (Equation~\ref{eq:degreeDist}), which is characteristic of scale-free networks \cite{Barabasi1999,Jeong2001}. On the other hand, if the equation describing the degree
distribution becomes exponential (Equation~\ref{eq:degreeExp}), the network is said to have an exponential degree distribution (such as the ER graph \cite{Bollobas1985}), indicating
the existence of some typical scales for degrees \cite{Amaral2000}. Our results show that the PPI networks of different species are grouped into two categories with scale-free
(\textit{M. musculus}, \textit{C. elegans}, \textit{D. melanogaster} and \textit{E. coli}) or exponential (\textit{S. cerevisiae}, \textit{H. sapiens} and \textit{A. thaliana}) degree
distributions (Table ~\ref{tab:exponents_static}). The above two properties, the scale-invariant property and the degree distribution, can be related through scaling theory in a
renormalization procedure \cite{Song2005}. At scale $\ell_B$, the degree of a hub $k$ changes to the degree of its box $k'$ (Equation~\ref{eq:degreeRG}). A new exponent $d_k$ relates
the fractal dimension $d_B$ and the scale-free exponent $\gamma$, which states the fact that $P(k)$ remains invariant under renormalization (Equation~\ref{eq:staticRelation}). The
corresponding values obtained were consistent with our theoretical predictions, confirming the validity of our approach (Tabel~\ref{tab:exponents_dynamic}).

{\bf The duplication-divergence model of network evolution enables the reconstruction of ancient interactomes}

\begin{figure}[!ht]
\begin{center}
\hbox{ A \includegraphics[width=5.5in]{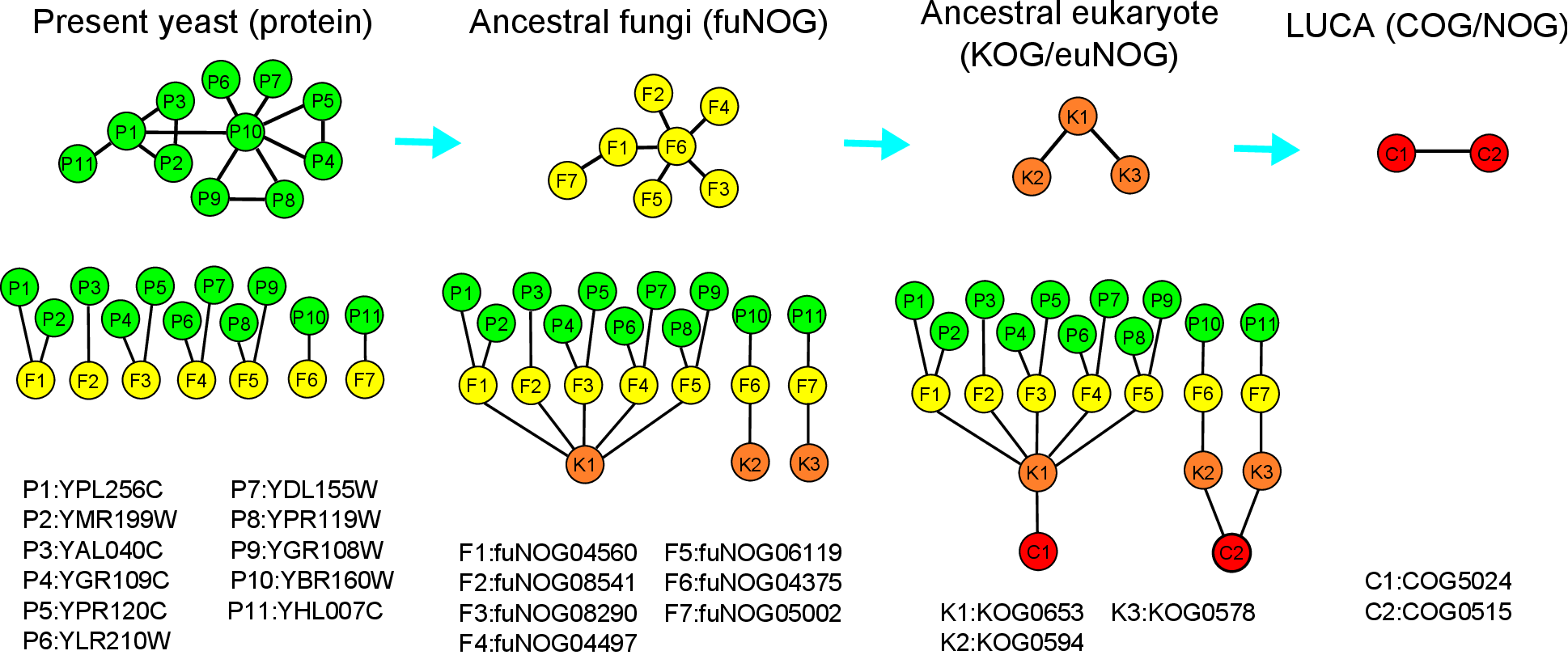}} \vspace{1cm}
 \hbox{B
\includegraphics[width=2.5in]{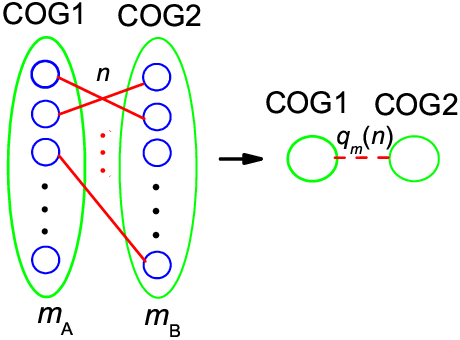} C
\includegraphics[width=2.5in]{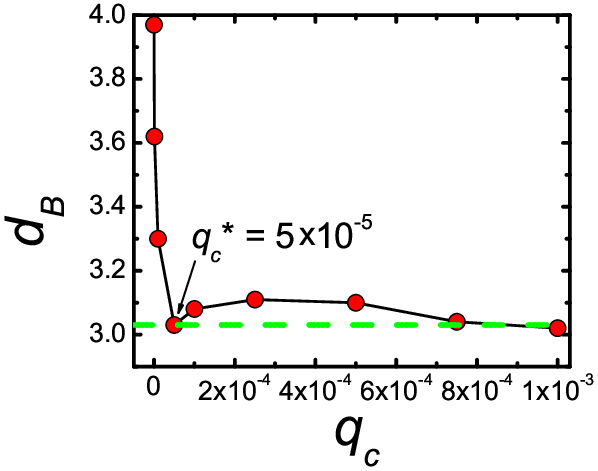}}
\end{center}
 \caption{{\bf An example of the reconstruction process of the \textit{S. cerevisiae} ancestral networks.} (A) Illustration of
the network reconstruction process. A subset of the empirical PPI network of \textit{S. cerevisiae} is shown. The phylogenetic trees demonstrate how the proteins are grouped into COGs
at different evolutionary levels. This information is used to identify the ancestral nodes. Note C2(COG0515) comprises other proteins which are not shown here. (B) The interaction
between each pair of COGs is assigned a probability $q_m(n)$ based on the duplication-divergence model. (C) The fractal dimension $d_B$ versus the cutoff $q_c$ for the ancestral
prokaryote network of yeast. By increasing $q_c$, $d_B$ approaches to the value of the present-day network (dashed line). We choose cutoff $q_c^* = 5 \times 10^{-5}$ so that the
ancestral network has the some fractal dimension as the present-day network. For $q_c>q_c^*$, $d_B$ remains (approximately) as a constant.} \label{fig:reconstruction}
\end{figure}

According to the duplication divergence model, present-day PPI networks evolved from ancestor PPI networks through protein duplication and loss events followed by diversification of
function and interactions. As the evolution of proteins can be well reconstructed using the concepts of orthology and paralogy, the {\bf C}lusters of {\bf O}rthologous {\bf
G}roups/{\bf N}onsupervised {\bf O}rthologous {\bf G}roups (COG/NOG) \cite{jensen_eggnog_2008} assignments of all proteins were retrieved from the eggNOG 2.0 database
\cite{Muller2010}. Recent proteins were assigned to the NOGs of the most recent level according to the lineage of the organism and the taxonomic resolution of eggNOG 2.0. If multiple
proteins were assigned to the same NOGs, duplication events have been reconstructed. This process was repeated between the NOG levels until the COG/NOG level, representing the last
universal common ancestor (LUCA), has been reached. The NOGs on the different (evolutionary) levels represent the ancestral proteins at this evolutionary timepoint. Figure \ref
{fig:reconstruction}A shows an example of the reconstruction process for a subset of the ancestral networks of \emph{S. cerevisiae}. The fuNOGs in Figure \ref {fig:reconstruction}A
(F1-F7) represent proteins in the ancestral fungi, KOGs/euNOGs (K1-K3) represent proteins in the ancestral eukaryotes and the COGs/NOGs (C1-C2) represent proteins
in the LUCA. The two yeast proteins P1 and P2 which are assigned to F1 indicate a duplication of F1 in \emph{S. cerevisiae}. \\

\begin{figure}[!ht]
\begin{center}
\includegraphics[width=5.5in]{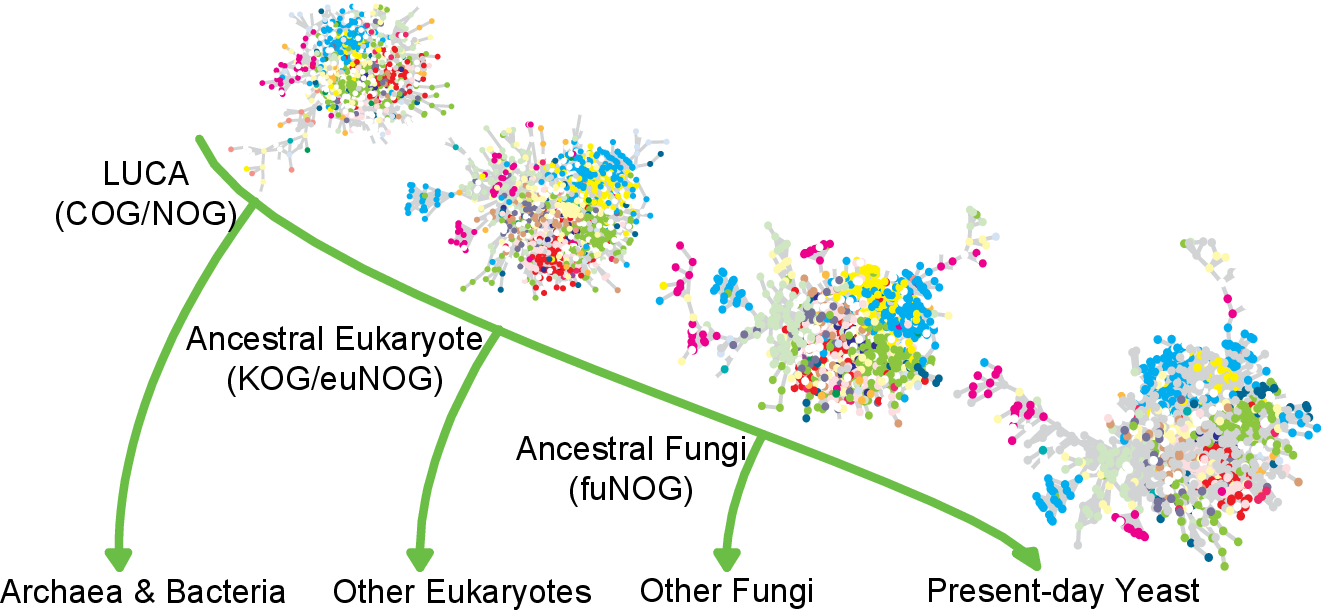}
\end{center}
\caption{{\bf Ancestral networks that were reconstructed for the
\textit{S. cerevisiae} PPI network.} Following the phylogenetic
tree, PPI networks on different evolutionary levels were
(re-)constructed: the present-day yeast (present-day protein), the
ancestral fungi (fuNOG, last common ancestor of fungi), the
ancestral eukaryote (KOG/euNOG, last common ancestor of animals,
plants and fungi), and the Last Universal Common Ancestor
(COG/NOG, last common ancestor of archaea, bacteria, and
eukaryotes). The colors of nodes represent the different
functional categories extracted from the eggNOG database
\cite{Muller2010}. } \label{fig:tree}
\end{figure}

While the ancestral nodes are obtained from the eggNOG database,
the reconstruction of ancestral interactions is much more difficult.
Although protein interactions are likely to be conserved between
pairs of orthologs (``interologs"), the limited knowledge about
recent interactions in many species and the link dynamics after
duplications make it impossible to use this principle for the
reconstruction of the links in ancient PPI networks. Thus, the
most promising approach is to transfer interactions measured in
today's PPI networks back in time, based on a model of link
evolution. Here we applied the duplication divergence model (see
Materials and Methods) to estimate the probability of the ancient
interactions based on today's PPI networks. A probability is assigned to the interaction
between each pair of COGs/NOGs (representing ancient proteins)
based on the number of possible
interactions between proteins in both COGs/NOGs and the number of
actually observed interactions in the present-day networks
(Figure~\ref{fig:reconstruction}B). The parameters required for
the model are derived by a fitting approach, so that the properties
of the resulting ancient networks resemble those of today's PPI
networks. We assume that general properties of
PPI networks are constant during evolution (Figure~\ref
{fig:reconstruction}C). The reconstruction is additionally constrained by
the underlying reconstruction of the ancient proteins. The
parameters defining which interactions are transferred back in
time are the fraction of interacting pairs in the ancestral
network at time $t$, $\alpha(t)$, the probability $p_x$ that an
interaction is retained after a duplication and the probability
$p_y$ that a new interaction is created \emph{de novo}. An
overview of the fitted parameters for all organisms is shown in
Table \ref{tab:parameters}. We observed that $p_x$ values range
between 0.5 and 0.7, but $p_y$ values are multiple orders of
magnitude smaller. These parameters indicate that link evolution
after duplication is the rule and de-novo creation is the
exception. The values are in good agreement with results from an
earlier study on \emph{S.
cerevisiae} \cite{Wagner2001}. A schematic representation of the reconstruction of the ancestral networks is given in Figure~\ref{fig:tree}, which shows the networks at the evolutionary levels that were reconstructed for \emph{S. cerevisiae}.\\

The consistency of the ancient PPI network was investigated by calculating their pair-wise overlaps. Therefore, the numbers of overlapping nodes and interactions between the organisms on all evolutionary levels were obtained (Figure \ref{fig:overlap_completeTable}). \emph{S. cerevisiae} has a relatively large overlap with all other species due to its network size, which is the largest of all organisms considered in the study. Whereas \emph{H. sapiens} shows relatively large overlaps with all other organisms, the highest overlap is, as expected, with \emph{M. musculus}, which is evolutionary most closely related to \emph{H. sapiens}. \emph{E. coli}, which has the third largest network of the organisms, exhibits small overlaps to all other organisms, except for \emph{S. cerevisiae}, which is the only other unicellular organism among the organisms of this study.\\

{\bf The change of interactome structures over time is explained by multiplicative growth mechanisms}

The reconstructed ancestral PPI network represent a series of
snapshots in the evolution of the present-day networks of the
respective species. By measuring the structural features of the
networks at these different time points, the growth principles of
the PPI network can be studied. Our results suggest a
multiplicative growth mechanism (see Materials and Methods) as
proposed in Ref.~\cite{Song2006}.

We first studied the PPI networks \emph{S. cerevisiae}, which is the largest network in our analysis. Figure~\ref{fig:dynamic_yeast}A shows that the time-dependent generator $n(\Delta
t)$, as well as the number of nodes $N(t)$ (see Equations~\ref{eq:exp_N} and~\ref{eq:n}), follows an exponential form with the nodes growth rate $r_N = 0.23(3)$/Gyr. The linear scaling
between $\ell(t_0)$ (the distance between two present-day proteins) and $\ell(t_\alpha)$ (the distance between two corresponding COGs/NOGs at time $t_\alpha$) on all evolutionary
levels is shown in Figure~\ref{fig:dynamic_yeast}B. The growth rate of the distances is found to be $r_l =  0.07(1)$/Gyr for the \emph{S. cerevisiae} network
(Figure~\ref{fig:dynamic_yeast}C). The two growth rates satisfy the condition $r_N/r_l = d_B$ (Figure~\ref{fig:dynamic_yeast}D and Tabel~\ref{tab:exponents_dynamic}). The result
relates the dynamic growth rates $r_N$ and $r_l$, to the static exponents $d_B$. This means that the nodes and distances do not grow independently but they grow at rates with a fixed
ratio which is equal to the fractal dimension $d_B$ and therefore conserve the fractal structure rather than becoming small-world. The linear scaling between $k(t_0)$ (the degree in
the present-day network) and $k(t_\alpha)$ (the degree of the corresponding COG/NOG at time $t_\alpha$) is shown in Figure~\ref{fig:dynamic_yeast}E. The growth rate for the
interactions $r_k \sim 0$ was found for \emph{S. cerevisiae}, which suggests $\gamma = \infty$ according to Equation~(\ref{eq:ns}). This implies that the \emph{S. cerevisiae} network
has an exponential degree distribution, which is consistent with the direct observation of the static network structures (Table~\ref{tab:exponents_dynamic} and
Figure~\ref{fig:degree}). While the multiplicative growth was originally proposed as a growth mechanism of nodes, distances and degrees \cite{Song2006}, simple generalization of the
same mechanism could be used to predict the growth rate of modularity (Equation~\ref{eq:m} and~\ref{eq:relation_dm}). For example, it was found that $d_M = 1.5(1)$ and $r_l =
0.07(1)$/Gyr,
Equation~(\ref{eq:relation_dm}) predicts $r_M = 0.11(2)$/Gyr. This assumes that the exponent $d_M$ is invariant, although the modules might involve with time. \\

\begin{figure}[!ht]
\begin{center}
\includegraphics[width=3in]{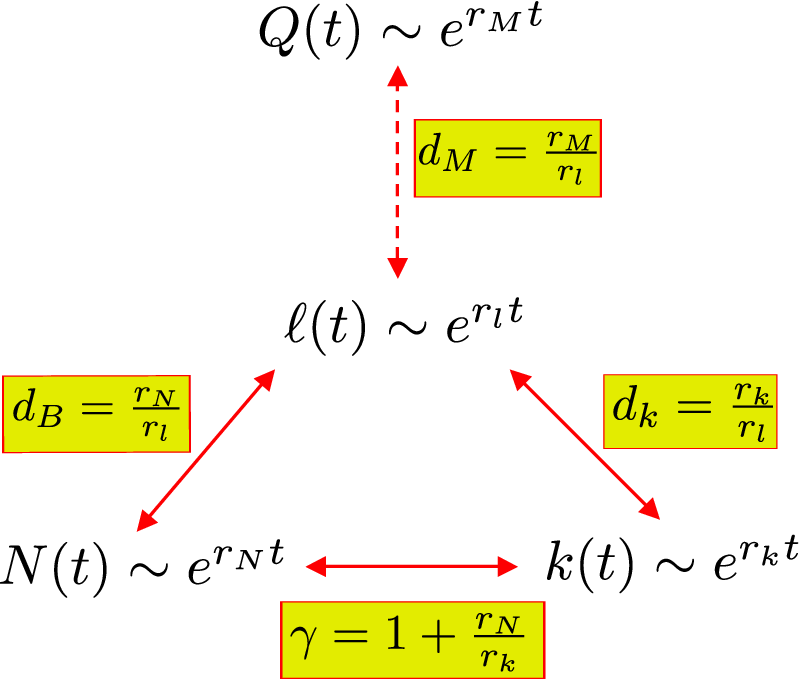}
\end{center}
\caption{{\bf Summary of the evolutionary mechanism.} Conservative
and multiplicative laws determine the static scaling exponents
($d_B$, $d_k$, $d_M$, $\gamma$) in terms of growth rates ($r_N$,
$r_l$, $r_k$, $r_M$). The three theoretical predictions ($d_B = r_N/r_l$, $\gamma = 1+r_N/r_k$, and $d_k = r_k/r_l$) have been corroborated by empirical calculations, while the remaining relation $d_M = r_M/r_l$ is a prediction open for test.}
\label{fig:relation}
\end{figure}

For studying the growth mechanisms in the PPI network of other
species, we selected the two further larger networks (\emph{E.
coli} and \emph{H. sapiens}) and one PPI network representing the
smaller networks (\emph{M. musculus}). We observed multiplicative
growth mechanisms also for these three PPI networks
(Table~\ref{tab:exponents_dynamic} and Figures
\ref{fig:dynamic_human}, \ref{fig:dynamic_mouse} and
\ref{fig:dynamic_ecoli}), indicating that these growth principles
are species-independent and thus universal. Furthermore, the
degree exponents, fractal dimensions and the modularities obtained
from this dynamic analysis were found in very good agreement with
those from the static analysis described above
(Table~\ref{tab:exponents_dynamic}). Our results confirm the
proposed relationship between the static scaling exponents and the
dynamic growth rates (Figure \ref{fig:relation}). The core of
the results are the exponential growth of the system quantities
($N$, $\ell$, $k$, $Q$), the relations between the static
exponents ($d_B$, $d_k$, $d_M$, $\gamma$) and the dynamic rates
($r_N$, $r_l$, $r_k$, $r_M$) (see Materials and Methods for a
detailed explanation).

\section*{Discussion}

The evolution of protein interaction networks is much less studied
compared to e.g. the evolution of DNA and aminoacid sequences.
This is not only a consequence of our sparse data on PPI networks, as
experimental approaches have intrinsic limitations and genome-wide
screens are very costly. Complete PPI networks, considering then entire
networks of protein-protein interactions across all possible
environmental conditions and developmental stages, are far from
being characterized even for unicellular model organisms such as
\emph{E. coli} or \emph{S. cerevisiae}. There are also a number of
conceptual questions how to study the evolution of networks. On
which levels are biological functions relevant for the evolution
of a PPI network (e.g. on the levels of binary interactions, protein
complexes, functional modules or entire networks)? How are the
emergent features of a PPI network selected in evolution (e.g. robustness
and stability)? How is the evolution of PPI networks connected with other
types of molecular networks? Most of these questions could hardly
be answered until now. Here we focus on one of the most basic
problems in PPI network evolution: what are the universal dynamic principles
by which PPI networks grow and change over time? The
increasing amount of PPI data for different organisms as well as
orthology reconstruction on different taxonomic levels allowed us
to investigate the network topology and growth of multiple
present-day and presumed ancient organisms in this study.\\

{\bf The structure of present-day PPI networks from multiple species}\\
Ideally, complete PPI networks from multiple species would have
been used for this study. Due to the limitations in the
experimental determination of PPI, no such data are so far available.
Therefore we had to compile a representative set of
input PPI networks from the heterogeneous, incomplete and
erroneous PPI data available. Although the integrative STRING
database very much simplified this task by providing the PPI data
from multiple organisms in a unified database scheme, the
distribution of experimental interaction scores was very different
among the selected species. This might result from different
experimental strategies, but makes the filtering by a static score
threshold questionable. For our study we expected the present-day
PPI networks to represent interactions of comparable strength and
confidence. A novel filtering approach based on the assumption of
self-similar topology was therefore implemented for the filtering
of the initial PPI data from the STRING database. We solved the
problem by applying a percolation analysis, which is based on the
idea of strength of links inspired from sociology, and has been
recently used to define functional brain networks from fMRI
signals \cite{Gallos2012A}. The percolation theory unambiguously
defines the critical threshold for the ranked scores in the STRING
database, which separates the small-world from the large-world of
self-similar structures: above or at the critical connectivity,
strong links form a highly modular, large-world fractal backbone,
and below the critical connectivity, weak ties establish shortcuts
between modules converting it to a small-world network
\cite{Granovetter1973, Gallos2012A}. The resulting score
thresholds varied significantly between the species. Considering
the scoring scheme of the STRING database, this might be explained
by varying proportions of individual vs. high-throughput
experiments in the database. However, in all networks a major
fraction of the interactions was removed through the filtering.
The remaining PPI are expected to form representative (as defined
by network topology) interaction networks on a species-specific
confidence level. Remarkably, a significant fraction of nodes was
removed as they were not represented on all taxonomic levels of
clusters of orthologous groups in the eggNOG database. This
phenomenon is not only present in the version 2.0 of this
database, but to a different extent also in the new version 3.0.
Besides technical reasons it might also be caused by complex
evolutionary histories (e.g. due to horizontal gene transfer) in
protein families. The filtered PPI networks in our study therefore
contain only proteins with a clearly traceable, mainly vertical
evolution. The success of the filtering operations can not be
directly assessed, as no additional gold-standard PPI data are
available. However we observed that structural and topological
properties of the filtered PPI networks were comparable also
beyond the initial assumption of self-similarity, indicating that
these data are a reasonable basis for further analysis in this
study.\\

{\bf Reconstructing ancient PPI networks based on the duplication-divergence model}\\
The duplication-divergence mechanism has been proposed by numerous
previous studies for the dynamic growth of PPI networks. Phenomena
like preferential attachment and correlation of evolutionary rate
vs. degree in PPI networks might be consequences of this growth
rules. To challenge this theory we developed an algorithm for the
reconstruction of ancient PPI networks based on present-day data.
Although the parameters of the duplication-divergence model might
be variable in evolutionary time, the limited data available make
only a general estimation possible. The duplication-divergence
model comprises two fundamental components: gene duplications and
link dynamics. The evolution of genes has been directly
reconstructed from clusters of orthologous groups. As these
clusters are widely used in bioinformatics e.g. for prediction of
gene function, the node structure of the ancient networks can be
considered to be very authentic. However, it embodies only a
fraction of the ancient proteomes. Proteins without present-day
interactions and proteins removed during the initial filtering are
missing, as well as proteins that have been lost in the evolution
of the species selected for this study. The ancient nodes
therefore specifically represent the ancestors of the nodes in the
present-day PPI networks.\\
Because the link dynamics are so far inaccessible by any orthology-driven approach, we developed an algorithm to reconstruct the most probable ancestral interactions based on the
stochastic duplication-divergence model. The fitting parameters in this model were determined from the COG data, which are independent of the network topology. As sequences of genes,
interactions are mainly created through gene duplication. However, previous studies did not agree whether it is more likely to retain or to lose an interaction after gene duplication
\cite{Wagner2001, presser_evolutionary_2008, gibson_questioning_2009}. In contrast to the evolution of sequences, \textit{de novo} gain of interactions are expected to occur much more
frequent than the \textit{de novo} formation of genes. This complicates the reconstruction of ancestral interactions significantly. Here we have developed a solution of this problem
based on a novel stochastic model of duplication/divergence constrained by the node structure (COG/NOG based) and the assumption of self-similar topology for the determination of the
interaction probability cutoffs. As expected, Table~\ref{tab:parameters} suggests for all species that the probability to retain an old interaction is equal or higher (0.5-0.7) than
that to lose an interaction, and is several orders higher than that to gain a new interaction (0.0001-0.0008). That is, $p_x > 1-p_x \gg p_y$. This means that the majority of
present-day interactions are inherited from ancestral interactions, while the generation of new interactions is much less frequent. A comparison of our results to values from earlier studies on \textit{S. cerevisiae} \cite{Wagner2001, presser_evolutionary_2008, gibson_questioning_2009} indicates very similar size ranges for the probability for retaining an interaction after a duplication and the probability for creating a new interaction \textit{de novo}. The good agreement between our results and results from earlier
studies, conducted on different datasets using different approaches, further supports the duplication divergence model of network
evolution.\\
While it is known that the duplication-divergence model results in an exponential growth of the network size \cite{Evlampiev2008}, there is no simple analytical way to predict the
dynamics of distance and modularity based on the model. However, it is important to note the connections between the network dynamics and the parameters in the duplication-divergence
model. For example, if $p_x =1.0$, the distances between proteins remain the same (Figure~\ref{fig:model}C) after duplications, while the number of proteins grows exponentially. This
results in a network of small-world structure and exponential dynamics, which shows that the duplication-divergence process does not necessary imply the fractality and the
multiplicative growth. When $p_x < 1.0$ as observed in Table~\ref{tab:parameters}, there is a probability that an old interaction is deleted, and the new protein is connected to the
old protein through a longer path (Figure~\ref{fig:model}C). This increases the distances between proteins. In fact, based on direct measurements of the reconstructed networks, we
found multiplicative (exponential) growth of distances. The multiplicative growth of both, nodes and distances, conserves the fractal/modular structure rather than becoming small-world.\\
A direct evaluation of the results is impossible as independent data on ancient PPI networks is unavailable. However, the consideration of different species in this study enables an
indirect assessment of our modeling results. Ideally, if the initial present-day PPI networks would be complete and free of errors, they should result in equivalent networks on the
ancient taxonomic levels. E.g, the present-day \emph{H. sapiens} and \emph{M. musculus} networks should predict the same ancient networks for the ancestral mammal, the ancestral
vertebrate etc. Assessing the pairwise similarities between the ancient PPI networks, we observed partial overlaps corresponding to the size of the present day networks (representing
completeness) and also according to the lifestyle and evolutionary distance of the organism. These results support the validity of the reconstruction algorithm based on the
duplication-divergence model, but they also indicate the substantial limitations of the
present-day PPI data.\\
Despite the strong evidence for the duplication-divergence model, the possibility of a model-dependent bias may still remain. The model favors a multiplicative growth rather than a linear growth over a relatively wide range of parameters. Further studies are required to test whether this preference is a biological consequence, or induced by the choice of the model. On the other hand, there exist other models \cite{Yang2008} consistent with a multiplicative growth. However, these models generally have no relevance to biological evolution, and therefore are not used in the study of PPI network evolution.\\

{\bf Universal dynamic principles determine the growth of PPI networks}\\
The explicit reconstruction of ancestral PPI networks for 7
selected species provides the unique opportunity to study their
growth dynamics. Although the filtering of initial PPI data and
the reconstruction algorithm utilize assumptions of fractal
topology, they do not necessarily result from multiplicative
growth. This means, whereas multiplicative growth implies fractal
topology, other growth mechanisms might produce fractal networks
as well, such as for instance a pure percolation process on the
network \cite{Bizhani2011}. Therefore we analyzed the growth of
number of nodes, number of edges, size and modularity of the
networks over time for the three larger networks and one selected
smaller network. In all networks we found a very good agreement
between the multiplicative growth principle and the observations in
the present-day and ancient PPI networks. Furthermore we found an
excellent matching between the results from static and dynamic
analysis, which are independent approaches. These results support
both the duplication-divergence model and multiplicative growth as
fundamental mechanisms in the long-term dynamics of PPI networks.

Our approach allowed to determine the network topologies of
multiple present-day and presumed ancient organisms based on two
widely used databases - STRING, providing information about
functional and physical protein interactions, and eggNOG,
providing information about the evolutionary relationships of
proteins. To our knowledge, such an extensive characterization of
multiple extant and ancient networks has not been performed until
now, as it is important for formulating and verifying
mathematical models describing the evolution of protein networks.
The network properties determined from topological network
analysis correspond well to the properties determined from dynamic
analysis based on the duplication-divergence evolutionary model.
This provides strong evidence for the correctness and the
universality of the proposed mathematical model of network
dynamics and evolution.

\section*{Materials and Methods}

{\bf Databases}

A database dump of the STRING database (release 8.3) was downloaded from ftp://string-db.org/ and a local database copy was set up. Binary protein interactions for the studied
organisms \cite{sayers_database_2012} (Table~\ref{tab:organisms}) with experimental scores above zero were extracted to obtain experimentally confirmed physical interactions. The
eggNOG database (release 2.0, ftp://eggnog.embl.de/eggNOG/2.0/) was used to obtain the assignment of proteins to clusters of orthologous groups (COGs/NOGs) on different taxonomic
levels. These levels are species-specific and defined in the eggNOG database. There are in total nine ancestral time levels for the organisms investigated: the ancestral primates
(prNOG), the ancestral rodents (roNOG), the ancestral mammals (maNOG), the ancestral vertebrates (veNOG), the ancestral insects (inNOG), the ancestral animals (meNOG), the ancestral
fungi (fuNOG), the ancestral eukaryotes (KOG/euNOG), and the LUCA (COG/NOG). Figure \ref{fig:tree} exemplifies the ancestral time levels for \textit{S. cerevisiae}. In the
initial filtering only proteins that were conserved on all evolutionary levels defined for the respective species were considered, thus every protein had an assignment to all its
evolutionary levels. Our reconstruction algorithm and reconstructed networks are
available at http://fileshare.csb.univie.ac.at/ppi\_evolution\_pone2013. \\

{\bf Reconstruction of the filtered present-day protein interaction networks}\\

The STRING confidence scores were used to assess the reliability
of the protein-protein interactions. For the identification of the
score threshold for reliable interactions the finding of Song et
al \cite{Song2005} that PPI networks are scale-invariant and
self-similar was taken as a basis. A threshold score $s_c^*$ above
which interactions were deemed reliable was determined and
confirmed for each organism by the following three independent
methods:\\

a) Percolation analysis. $s_c^*$ can be found as the threshold of
a percolation transition of the network. When networks are
reconstructed for all possible confidence scores, the percolation
threshold $s_c^*$ represents the first jump in the size of the
largest cluster, while the size of the second largest cluster
peaks at this point (see Figure~\ref{fig:percolation}A). The
percolated cluster, also called giant connected component, is
formed by links whose confidence score is higher or equal to
$s_c^*$. We observed a series of jumps in the percolation process,
which suggests a multiplicity of percolation transitions
\cite{Gallos2012A, Gallos2012C}. This is different from a random
percolation (Figure~\ref{fig:percolation}A inset), where only
single transition point exists. Our results show that the
percolation process of PPI networks is more complicated than a
simple uncorrelated percolation process, due to the modular
organization and the strong correlations between protein
interactions.\\

b) MEMB-algorithm. The box-covering algorithm MEMB \cite{Song2007} (Figure~\ref{fig:percolation}B) was used to tile the network with the minimum number of boxes $N_B$ of a given box diameter $\ell_B$. $\ell_B$ was defined such that the maximum distance in a box is smaller than $\ell_B$, and distance was measured as the number of links on the shortest path between two proteins. A power-law scaling of $N_B$ and $\ell_B$ at $s_c^*$ confirms the fractality of the network at the percolation threshold (Figure~\ref{fig:percolation}C).\\

c) Renormalization group analysis. The renormalization group
approach \cite{Rezenfeld2010} was used for another confirmation of
the $s_c^*$ threshold as the transition point between small-world
and fractal phases. The renormalized network is built by replacing
the boxes by ``supernodes" and two supernodes are connected if
there is at least one link between two nodes in their respective
boxes. The relationship between the average degree of the
renormalized network, $z_B$, and the average number of nodes in
each box $x_B = N/N_B = \ell_B^{d_B}$ gives information about
whether the network is small-world (positive slope), fractal
(negative slope) or at the phase transition $s_c^*$ (slope of 0)
(see Figure~\ref{fig:percolation}D). \\

The addition of links of scores below $s_c^*$ (defined from
percolation analysis, Figure~\ref{fig:percolation}A) converts a
fractal network (above $s_c^*$) into a small-world network. That
is, the power-law relation (Equation~\ref{eq:fractal}) transforms
into an exponential decay characteristic of small-world
(MEMB-algorithm, Figure~\ref{fig:percolation}C), and the slopes
become positive in Figure~\ref{fig:percolation}D (renormalization
group analysis). Therefore, the three independent methods are
consistent with each other. From the resulting networks, the
largest connected component at $s_c$* was used for topological
analysis. \\

{\bf Topological properties of the networks}\\
The fractal dimension $d_B$ was measured from the MEMB algorithm,
by fitting the relationship between the minimum number of boxes
$N_B$ and the box diameter $\ell_B$ to a power-law function
\cite{Song2005} (see Figure~\ref{fig:percolation}C for \textit{S.
cerevisiae} and \textit{M. musculus}):
\begin{equation}
N_B(\ell_B) \sim \ell_B ^ {-d_B}, \label{eq:fractal}
\end{equation}
where $d_B$ is the fractal dimension which characterizes the
self-similarity between different topological scales of the
network. The values of $d_B$ for all species are summarized in
Table~\ref{tab:exponents_static}.

The degree distribution $P(k)$ was measured and the degree
exponents $\gamma$ \cite{Barabasi1999} were determined. For some
networks (\textit{M. musculus}, \textit{C. elegans}, \textit{D.
melanogaster} and \textit{E. coli}) it was shown to follow a power
law distribution with degree exponent $\gamma$:
\begin{equation}
P(k) \sim (k+k_0)^{-\gamma}, \label{eq:degreeDist}
\end{equation}
where $k_0$ is a small cutoff degree. For others (\textit{S.
cerevisiae}, \textit{H. sapiens} and \textit{A. thaliana}) the
parameters became $\gamma \rightarrow \infty$, $k_0 \rightarrow
\infty$ with fixed $k_c =k_0/\gamma$ and the equation had an
exponential form:
\begin{equation}
P(k) \sim e^{-k/k_c}, \label{eq:degreeExp}
\end{equation}
Figure~\ref{fig:degree} shows $P(k)$ of two species, \textit{S.
cerevisiae} (exponential) and \textit{M. musculus} (scale-free),
which are characteristic of the behaviors found across all
species. Table~\ref{tab:exponents_static} summarizes the values of
$\gamma$ for all the species.

The above two properties, the scale-invariant property,
Equation~(\ref{eq:fractal}), and the degree distribution,
Equation~(\ref{eq:degreeDist}), can be related through scaling
theory in a renormalization procedure\cite{Song2005}. At scale
$\ell_B$, the degree of a hub $k$ changes to the degree of its box
$k'$, through the relation:
\begin{equation}
k' = \kappa(\ell_B)k, ~~~\mbox{with    } \kappa(\ell_B)\sim
\ell_B^{-d_k}, \label{eq:degreeRG}
\end{equation}
A new exponent $d_k$ relates the fractal dimension $d_B$ and the
scale-free exponent $\gamma$ through
\begin{equation}
\gamma = 1 + d_B/d_k, \label{eq:staticRelation}
\end{equation}
which states the fact that $P(k)$ remains invariant under renormalization. For the \textit{S. cerevisiae} PPI network, we found $\gamma \sim \infty$, $d_B = 3.0(2)$, and $d_k \sim 0$,
and for the \textit{M. musculus} PPI network, we found $\gamma = 2.9(1)$, $d_B = 1.7(1)$, and
$d_k = 0.8(1)$ (Figure~\ref{fig:dk}). The values of $d_k$ are summarized in Table~\ref{tab:exponents_dynamic}. The results are consistent with our theoretical prediction, Equation~(\ref{eq:staticRelation}).\\

{\bf Modularity}\\
The modular organization \cite{Gallos2012A,Galvao2010,Girvan2002}
of the network was investigated by the analysis of the links
inside and between topological modules. Modules were defined by
the boxes detected by MEMB algorithm. To capture the degree of
modularity of the network, the modularity ratio ${Q}(\ell_B)$ was
defined as a function of the size of the modules, $\ell_B$:
\begin{equation}
{Q}(\ell_B) = \frac{1}{N_c} \sum_{i=1}^{N_c} \frac{L^i_{\rm
in}}{L^i_{\rm out}}, \label{eq:mod}
\end{equation}
where $L^i_{\rm in}$ is the number of links between nodes inside
the module $i$, $L^i_{\rm out}$ is the number of links from module
$i$ connecting to other  modules and $N_c$ is the number of
modules needed to tile the network for given size $\ell_B$. Large
values of $Q$ correspond to a structure where the modules are well
separated and therefore to a higher degree of modularity. The
degree of modularity depends on the scale as:
\begin{equation}
{Q}(\ell_B) \sim \ell_B^{d_M} \label{eq:dm}
\end{equation}
which defines the modularity exponent (see Figure~\ref{fig:modularity}). \\

{\bf Construction of the ancient protein interaction networks}\\
The reconstruction of the ancient networks is based upon two integral parts: the identification of the ancestral proteins due to their evolutionary relationships and their assignment
to COGs/NOGs (described above) and a duplication-divergence model describing the link dynamics during evolution. A fundamental assumption for both parts is that the structural network
features are time-invariant.\\

The ancestral nodes were obtained from the assignment of present-day proteins to COGs/NOGs provided by the eggNOG database on different time levels.\\

The next crucial step was to decide when to transfer present-day interactions to the presumptive ancient network. Each COG could comprise several proteins, and the proteins in the same
COG pair may or may not interact. Rather than transferring every present-day interaction, it is necessary to assess the probability that the respective COGs interact. For example, if
two COGs comprise 10 proteins each, but there is only one interaction (out of 100 total possible interactions) between these proteins in the present-day network, it is improbable that
these COGs (or the ancient proteins they represent) actually interacted.\\

In order to estimate this probability, the relationship between
the number of total possible interactions and the number of actual
interactions between the proteins which participate in these COGs
is considered. As illustrated in Figure~\ref{fig:reconstruction}B,
if two COGs A and B comprise $m_A$ and $m_B$ proteins each, then
there are $m = m_A \times m_B$ total possible interactions between
the proteins in the COGs. Out of the $m$ possible interactions,
let $n$ be the number of interactions that are actually detected
in the present-day experimental data. One simple way is to assume
the ancestral link probability between COGs A an B is proportional
to $n/m$. However, this assumption is oversimplified, since this
probability does not only depend on the ratio $n/m$, but also on
the value of $m$. For example, depending on the data it is 10
times more probable to find $n=1$ actual interaction out of $m=2$
total possible ones, than to find $n=4$ actual interactions out of
$m = 8$ possible ones, although they have equal ratio $n/m$.\\

In the reconstruction method, a probability $q_m(n)$ (see below
how $q_m(n)$ is calculated) is assigned to the ancestral
interaction between the two COGs. The value of $q_m(n)$ is
calculated from a stochastic model described below. This way, a
network of COG-COG interactions with weighted edges given by
$q_m(n)$ is constructed, where the edges with large weights are
regarded as the most-likely
interactions constituting the ancestral network.\\

The final step is to determine a proper cutoff of $q_m(n)$ since
COG pairs with low $q_m(n)$ would most probably not interact. Only
interactions with probability higher than $q_c$ ($q_m(n) > q_c$)
are included in the analysis. Changing this cutoff value allows to
switch the sensitivity or selectivity of the ancestral
interactions. To determine the cutoff, it is required that the
reconstructed networks at different time levels have invariant
topological features. In practice, the fractal dimension $d_B$ in
each ancestral network is measured explicitly as a function of the
cutoff $q_c$ (Figure~\ref{fig:reconstruction}C), and a critical
value of $q_c^*$ is determined when $d_B$ reaches to the same
value as the present network. For example, in the case of the
\textit{S. cerevisiae}, we find $q_c^* = 5\times10^{-5}$.
\\

In order to estimate the probability of the ancestral interactions $q_m(n)$, we developed a symmetric stochastic evolution model of the protein interaction network based on
duplication-divergence processes \cite{Ohno1970, Li1997, Patthy1999, Taylor2004}. The model takes into account the deletion of duplication-derived interactions and \textit{de novo}
creation of interactions. An analytical function of link probability is derived to compare with
experimental data and determine the parameters.\\

Based on the mechanism of genomic duplication and divergence two general modes are considered: (i) Mode I (Figure~\ref{fig:model}A): protein A initially interacts with protein B, and protein A is duplicated into two proteins A and A'. The duplicated proteins A and A' have equal probability $p_x$ to copy the interaction link with protein B. (ii) Mode II (Figure~\ref{fig:model}B): protein A and B do not interact with each other initially. There is a probability $p_y$ that the duplicated proteins A or A' gains a new interaction with protein B.\\
The evolution of the network is completely specified by the
parameters $p_x$, $p_y$ and its initial condition. $p_i$ describes
the probability of an interaction between any pair of new proteins
after $i$ total duplications (protein A and B duplicates $i_A$ and
$i_B$ times each, and $i = i_A + i_B$). Two successive duplication
steps can be represented by the recursive relation of $p_i$
\begin{equation}
p_i = p_{i-1}p_x + (1-p_{i-1})p_y, \label{eq:recursion}
\end{equation}
where the first term comes from the contribution of the existing
link at $(i-1)^{\rm th}$ step, and the second term is from the
non-existing link. Equation~(\ref{eq:recursion}) can be solved
recursively, producing a formula of $p_i$ which only depends on
$p_x$, $p_y$ and the initial condition:
\begin{equation}
p_i (p_0, p_x, p_y) = p_0 \eta ^ i + \frac{1-\eta^i}{1-\eta}p_y,
\label{eq:pi}
\end{equation}
where $\eta \equiv p_x - p_y$. Here $p_0$ describes the initial condition: $p_0 = 1$ if the pair of proteins initially interact with each other, otherwise, $p_0 = 0$.\\

After $i_A$ ($i_B$) duplications, the initial protein A (B) evolves into a cluster comprising $m_A = 2^{i_A}$ ($m_B = 2^{i_B}$) present-day proteins. $m = m_A \times m_B$ is the total
number of possible interactions, and $i = \log_2(m)$ is the total number of duplications (Figure~\ref{fig:model}B). Let $p_i(1) \equiv p_i(p_0 = 1, p_x, p_y)$, and $p_i(0) \equiv
p_i(p_0 = 0, p_x, p_y)$. For a pair of clusters with $m$ total possible interactions, the probability $p_m(n)$ that $n$ pairs of these proteins actually interact, given that each pair
have independent probability $p_i$, is represented by a binomial distribution. If the initial pair of ancestral proteins interact, then $p_m(n) =
\binom{m}{n}p_{i}^n(1)[1-p_{i}(1)]^{m-n}$; if they do not initially interact, then $p_m(n) = \binom{m}{n}p_{i}^n(0)[1-p_{i}(0)]^{m-n}$. $p_m(n)$ of a network is a combination of these
two cases. Assume that $\alpha(t)$ is the fraction of interacting pairs out of total possible pairs in the ancestral network at time $t$. $p_m(n)$ can be calculated as:
\begin{equation}
p_m(n)=   \alpha(t) \binom{m}{n}p_{i}^n(1)[1-p_{i}(1)]^{m-n} +
[1-\alpha(t)] \binom{m}{n}p_{i}^n(0)[1-p_i(0)]^{m-n},
\label{eq:pmn}
\end{equation}
The first term describes the interacting pairs in the ancestral network, and the second term is from the non-interacting pairs. Note that $p_m(n)$ depends on time $t$ since we assumed
that $\alpha(t)$ could be different at different time levels. \\

Equation~(\ref{eq:pmn}) depends on three parameters $p_x$, $p_y$ and $\alpha(t)$ for each time $t$. It was assumed that $p_x$ and $p_y$ are constants at different time levels, and
$\alpha(t)$ is time-dependent. To determine these parameters, $p_m(n)$ is fitted to the values derived from the present-day networks and COG data. For each evolutionary level $t$, we
first found the number of possible COG pairs that contains $m$ total possible interactions, $N_{t,m}$. Out of $N_{t,m}$ total pairs, we counted the number of COG pairs that have $n$
actual interactions, $N_{t,m,n}$.

Statistically, the ratio $N_{t,m,n}/N_{t,m}$ should represent the probability $p_m(n)$. In order to find the best fitting, we minimized the objective function
\begin{equation}
F = \sum_{t=1}^{t_{\rm max}} \sum_{m=2}^{m_{\rm max}} \sum_{n=0}^{m}\left[\log \left( p_m(n) \right) -\log \left( \frac{N_{t,m,n}}{N_{t,m}} \right) \right]^2, \label{eq:fit}
\end{equation}
where $t_{\rm max}$ is the maximum time level, and $m_{\rm max}$ is the maximum $m$ used in fitting. Our objective function is very similar to the standard residual sum of squares
(RSS). The logarithm values are used here because $p_m(n)$ has an exponential behavior (Figure~\ref{fig:test}). Minimization of Equation~(\ref{eq:fit}) is an unconstrained nonlinear
optimization problem on multiple parameters, which was handled by the function {\it fminsearch} in MATLAB R2012a.

$p_m(n)$ was fitted to the measured values for all organisms. To have meaningful sample sizes, $m$ was restricted to be between 2 and 8. Figure~\ref{fig:test} shows the results of
three species: \textit{S. cerevisiae}, \textit{M. musculus}, and \textit{H. sapiens}. The fitted curves are in good agreement with empirical data. The fitted parameters for all species are summarized in Table~\ref{tab:parameters}.\\

Since $q_m(n)$ is the probability to have an ancestral link for a given $m$ and $n$, it is proportional to $\alpha(t) \binom{m}{n}p_{i}^n(1-p_{i})^{m-n}$, which is the first term in
Equation~(\ref{eq:pmn}). With a proper normalization, we obtained:
\begin{equation}
q_m(n) = \frac{\alpha(t) \binom{m}{n}p_{i}^n(1)[1-p_{i}(1)]^{m-n}
}{\alpha(t) \binom{m}{n}p_{i}^n(1)[1-p_{i}(1)]^{m-n} +
[1-\alpha(t)] \binom{m}{n}p_{i}^n(0)[1-p_i(0)]^{m-n}}.
\label{eq:qmn}
\end{equation}
Equation~(\ref{eq:qmn}) was used to reconstruct the ancestral networks (see Figure~\ref{fig:reconstruction}B) with fitted parameters from Table~\ref{tab:parameters}.\\

{\bf Determination of the growth principles}\\
To determine the dynamical processes governing the changes in network structures over time, the growth rates of nodes, distances and degrees were empirically determined. In detail, the following values were determined directly from the networks at each timepoint $t$: the number of nodes $N$, the number of links $k$, the distance $\ell$ between two COGs. Our results support the multiplicative mechanism proposed in \cite{Song2006} to account for the fractal, modular and scale-free nature of PPI network structures.\\

The determined growth rates were set in relation to the scaling
exponents of the networks, which were obtained from the static
topological network analysis. Estimations for the divergence times between the organisms were derived from \cite{hedges_timetree_2006} and are listed in Table~\ref{tab:divergenceTimes}, which provide the time $t_\alpha$ representing the time levels of COGs/NOGs.\\

\begin{figure}[!ht]
\hbox{ A \includegraphics[width=2.5in]{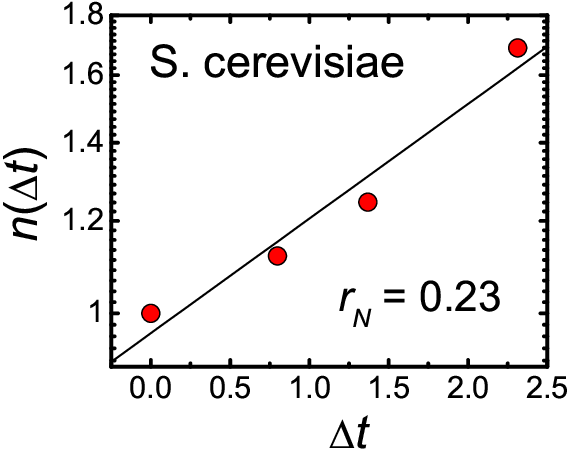}
       B \includegraphics[width=2.5in]{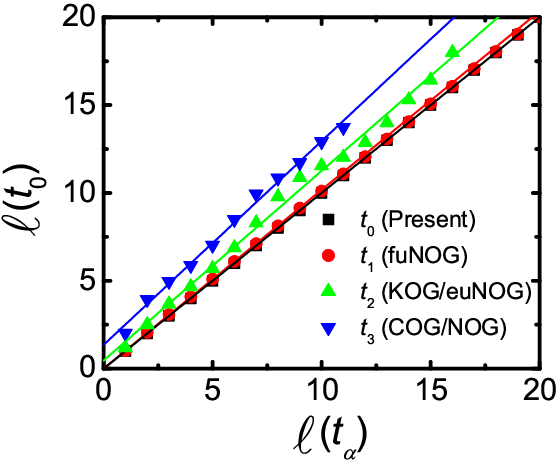}}
  \vspace{.5cm}
\hbox{ C \includegraphics[width=2.5in]{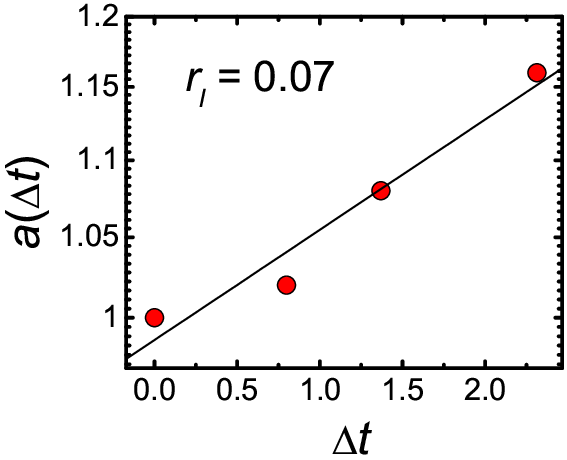}
       D \includegraphics[width=2.5in]{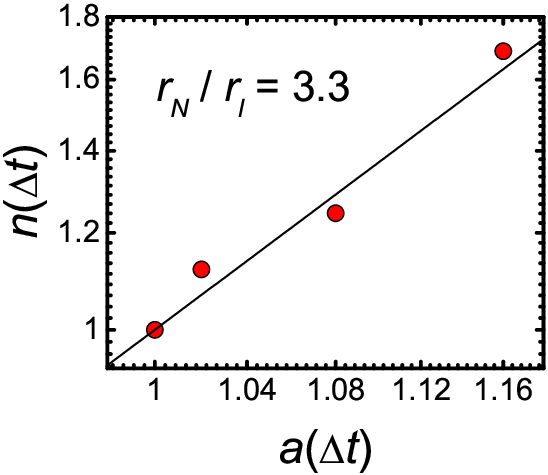}}
  \vspace{.5cm}
\hbox{ E
\includegraphics[width=2.5in]{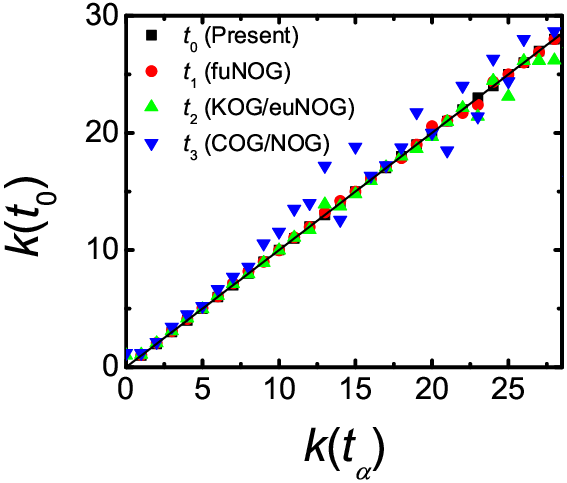}}
\caption{{\bf Multiplicative growth mechanism of the \emph{S. cerevisiae} PPI network.} (A) Semi-log plot of $n(\Delta t)$ vs. $\Delta t$. The growth rate $r_N = 0.23(3)$ is obtained
from a linear fitting. The unit of time is Gyr. (B) Scaling between $\ell(t_0)$ and $\ell(t_\alpha)$. Each point is an average over many pairs of nodes in the network with the same
$\ell(t_\alpha)$. The slope of the linear fitting gives $a(\Delta t)$, where $\Delta t = t_0 - t_\alpha$ is the time difference between two evolutionary levels. (C) Semi-log plot of
$a(\Delta t)$ vs. $\Delta t$. The growth rate $r_l = 0.07(1)$ is obtained from a linear fitting. (D) Log-log plot of $n(\Delta t)$ vs. a($\Delta t$). The scaling shows that the ratio
between two growth rates ($r_N/r_l = 3.3(8)$), is close to the static measure of the fractal dimension $d_B = 3.0(2)$. This confirms the relationship Equation (\ref{eq:relation_dB}).
(E) Scaling between $k(t_0)$ and $k(t_\alpha)$. Each point is an average over many nodes with the same $k(t_\alpha)$. Large degrees ($k>27$) are not included in this plot since there
is not enough number of samples to make meaningful statistics. The slope of the linear fitting gives $s(\Delta t) \sim 1.0$, which is consistent with an exponential degree
distribution.} \label{fig:dynamic_yeast}
\end{figure}

The increase in the number of nodes over time is best approximated
by an exponential function:
\begin{equation}
N(t) \sim e^{r_N  t}, \label{eq:exp_N}
\end{equation}
with a growth rate of the number of nodes $r_N$. This implies the
multiplicative growth form of $N$ with a time-dependent generator
$n(\Delta t)$:
\begin{equation}
N(t) = n(\Delta t) ~ N(t_\alpha), ~~~\mbox{with    } n(\Delta t) =
e^{r_N  \Delta t}, \label{eq:n}
\end{equation}
where $\Delta t = t - t_\alpha$. Figure~\ref{fig:dynamic_yeast}A and Figure~\ref{fig:dynamic_human}A show this growth mechanism for \textit{S. cerevisiae} and \textit{H. sapiens}. Table~\ref{tab:exponents_dynamic} summarizes measured $r_N$ of all species.

Next, we consider the distance between two COGs in an ancestral
network, $\ell(t_\alpha)$, and compare with the corresponding
distance $\ell(t_0)$ in the present network. $\ell(t_0)$ is
measured as the distance between the two hubs in each COG, where a
hub is the protein with maximum degree inside each COG. If two
hubs have the same degree, then the average value was taken. The
evolution of distance $\ell$ can be modeled by a similar form:
\begin{equation} \ell(t) = a(\Delta t) ~ \ell(t_\alpha),
~~~\mbox{with    } a(\Delta t) = e^{r_l\Delta t}, \label{eq:a}
\end{equation} This suggests an exponential growth of distances
instead of a linear growth. The multiplicative growth of $N$ and
$\ell$ is consistent with the fractal scaling law
Equation~(\ref{eq:fractal}). On the contrary, a combination of
exponential growth of nodes and linear growth of distances would
result in an exponential scaling between nodes and distances,
which represents a small-world network \cite{Song2006}.
Figures~\ref{fig:dynamic_yeast}B,~\ref{fig:dynamic_human}B,~\ref{fig:dynamic_mouse}A,
and~\ref{fig:dynamic_ecoli}A show the linear scalings between
$\ell(t_0)$ ($t_0$ is the present time) and $\ell(t_\alpha)$ for
four representative species, \textit{S. cerevisiae}, \textit{H.
sapiens}, \textit{M. musculus}, and \textit{E. coli}. $a(\Delta
t)$ was obtained by liner fittings and was used to calculate the
growth rates $r_l$ (see Figure~\ref{fig:dynamic_yeast}C for
\textit{S. cerevisiae} and Figure~\ref{fig:dynamic_human}C for
\textit{H. sapiens}). The values of $r_l$ of all species are
listed in Table~\ref{tab:exponents_dynamic}.

The growth Equations (\ref{eq:n}) and (\ref{eq:a}) can be combined
to obtain a power-law relation between the distances and the
number of proteins with an exponent $d_B$ given by the ratio of
the growth rates, \begin{equation} d_B = \frac{\ln n(\Delta
t)}{\ln a(\Delta t)} = \frac{r_N}{r_l}, \label{eq:relation_dB}
\end{equation}\\ Equation~(\ref{eq:relation_dB}) shows the
relation between the static exponent $d_B$ and dynamic growth
rates $r_N$ and $r_l$. This theoretical prediction is tested in
Figures~\ref{fig:dynamic_yeast}D,~\ref{fig:dynamic_human}D,~\ref{fig:dynamic_mouse}B,
and~\ref{fig:dynamic_ecoli}B, which confirm a power-law relation
between $n(\Delta t)$ and $a(\Delta t)$.
Table~\ref{tab:exponents_dynamic} shows that $d_B$ measured from
static network structure is in good agreement with the value
$r_N/r_l$ predicted from dynamic growth rates.

The number of interactions $k(t_\alpha)$ of each COG at time
$t_\alpha$ was compared with the degree $k(t_0)$ in the present
yeast network, where $k(t_0)$ was the degree of the hub in each
COG. Our results
(Figures~\ref{fig:dynamic_yeast}E,~\ref{fig:dynamic_human}E,~\ref{fig:dynamic_mouse}C,
and~\ref{fig:dynamic_ecoli}C) show that the number of interactions
$k$ also follows a general form of multiplicative growth with a
time-independent generator $s(\Delta t)$:
\begin{equation}
k(t) = s(\Delta t) ~ k(t_\alpha), ~~~\mbox{with    }s(\Delta t) =
e^{r_k \Delta t} \label{eq:s}
\end{equation}\\
$s(\Delta t)$ was measured from linear fitting of this scaling between $k(t_0)$ and $k(t_\alpha)$. The growth rates $r_k$ were measured and listed in Table~\ref{tab:exponents_dynamic}.
In particular, for networks of exponential degree distributions (such as \textit{S. cerevisiae}, \textit{H. sapiens} and \textit{A. thaliana}), $s(\Delta t) \sim 1.0$  and $r_k \sim 0$
(see Figure~\ref{fig:dynamic_yeast}E for \textit{S. cerevisiae} and Figure~\ref{fig:dynamic_human}E for \textit{H. sapiens}), which suggests that the degrees are invariant.

This dynamic behavior of degrees is consistent with the static
measure of the degree distribution. Using the density conservation
law of degree distribution over evolution \begin{equation}
N(t_\alpha) P(k(t_\alpha)) dk(t_\alpha) = N(t) P(k(t)) dk(t)
\end{equation} the degree distribution
Equation~(\ref{eq:degreeDist}), and the growth laws Equations~(\ref{eq:n}) and (\ref{eq:s}), the following relationship between the static exponent $\gamma$ and the dynamic rates $r_N$
and $r_k$ was obtained: \begin{equation} \gamma = 1 + \frac{\ln n(\Delta t)}{\ln s(\Delta t)} = 1+\frac{r_N}{r_k}. \label{eq:ns} \end{equation}\\ Equation~(\ref{eq:ns}) was tested in
Figures~\ref{fig:dynamic_mouse}D and \ref{fig:dynamic_ecoli}D for scale-free networks (such as \textit{M. musculus}, \textit{C. elegans}, \textit{D. melanogaster} and \textit{E.
coli}). For exponential networks (such as \textit{S. cerevisiae}, \textit{H. sapiens} and \textit{A. thaliana}), Equation~(\ref{eq:ns}) suggests $\gamma \sim \infty$ since $r_k \sim 0$
as measured in Figure~\ref{fig:dynamic_yeast}E  and Figure~\ref{fig:dynamic_human}E. The comparison between $\gamma$ and $1+\frac{r_N}{r_k}$ is shown in
Table~\ref{tab:exponents_dynamic} with good agreements.

The relationship between $N$, $\ell$ and $k$ is closed by the
third equation:
\begin{equation}
d_k = \frac{\ln s(\Delta t)}{\ln a(\Delta t)} = \frac{r_k}{r_l}.
\label{eq:sa}
\end{equation}\\
This was tested in Figures~\ref{fig:dynamic_mouse}E and \ref{fig:dynamic_ecoli}E for scale-free networks. For exponential networks, we found $r_k \sim 0$, and therefore $d_k \sim 0$,
which agrees with the static measurement (Table~\ref{tab:exponents_dynamic}).

Equations~(\ref{eq:relation_dB}), (\ref{eq:ns}), and (\ref{eq:sa})
relate the static exponents $d_B$, $\gamma$, and $d_k$ to the
dynamic growth rates $r_N$, $r_l$, and $r_k$. Combining the three
equations together, the static relationship
Equation~(\ref{eq:staticRelation}) is recovered, which is
originally derived from scaling argument \cite{Song2005}.

Similar to the growth laws of $N$, $\ell$ and $k$, an exponential
growth of $Q$ is assumed:
\begin{equation}
Q(t) =  e^{r_M t}, \label{eq:m}
\end{equation}
and a relationship is predicted as:
\begin{equation}
d_M = \frac{r_M}{r_l}, \label{eq:relation_dm}
\end{equation}
This assumes that the modularity exponent $d_M$ is invariant during evolution. Direct test of this assumption would require detailed analysis of network structure and protein functions, which was left for future study.

The above results are summarized in Figure~\ref{fig:relation}. At
the core of the results is the exponential growth of the system
quantities ($N$, $\ell$, $k$, $M$), and the relations between the
static exponents ($d_B$, $d_k$, $d_M$, $\gamma$) and dynamic rates
($r_N$, $r_l$, $r_k$, $r_M$). Therefore, the multiplicative growth
provides a fundamental mechanism for the evolutionary
principle of PPI networks.

\section*{Acknowledgments}
We thank Her\'{n}an D. Rozenfeld, Lazaros K. Gallos,
and Chaoming Song for valuable discussions.

\bibliography{short_resub}

\begin{thebibliography}{10}
\providecommand{\url}[1]{\texttt{#1}}
\providecommand{\urlprefix}{URL }
\expandafter\ifx\csname urlstyle\endcsname\relax
  \providecommand{\doi}[1]{doi:\discretionary{}{}{}#1}\else
  \providecommand{\doi}{doi:\discretionary{}{}{}\begingroup
  \urlstyle{rm}\Url}\fi
\providecommand{\bibAnnoteFile}[1]{%
  \IfFileExists{#1}{\begin{quotation}\noindent\textsc{Key:} #1\\
  \textsc{Annotation:}\ \input{#1}\end{quotation}}{}}
\providecommand{\bibAnnote}[2]{%
  \begin{quotation}\noindent\textsc{Key:} #1\\
  \textsc{Annotation:}\ #2\end{quotation}}
\providecommand{\eprint}[2][]{\url{#2}}

\bibitem{mika_protein-protein_2006}
Mika S, Rost B (2006) Protein-protein interactions more conserved within
  species than across species.
\newblock PLoS Comput Biol 2: e79.
\bibAnnoteFile{mika_protein-protein_2006}

\bibitem{zinman_biological_2011}
Zinman GE, Zhong S, {Bar-Joseph} Z (2011) Biological interaction networks are
  conserved at the module level.
\newblock BMC Syst Biol 5: 134.
\bibAnnoteFile{zinman_biological_2011}

\bibitem{gibson_improving_2011}
Gibson TA, Goldberg DS (2011) Improving evolutionary models of protein
  interaction networks.
\newblock Bioinformatics 27: 376--382.
\bibAnnoteFile{gibson_improving_2011}

\bibitem{Fields2005}
Fields S (2005) High-throughput two-hybrid analysis.
\newblock FEBS J 272: 5391--5399.
\bibAnnoteFile{Fields2005}

\bibitem{suter_two-hybrid_2008}
Suter B, Kittanakom S, Stagljar I (2008) Two-hybrid technologies in proteomics
  research.
\newblock Curr Opin Biotechnol 19: 316--323.
\bibAnnoteFile{suter_two-hybrid_2008}

\bibitem{koegl_improving_2007}
Koegl M, Uetz P (2007) Improving yeast two-hybrid screening systems.
\newblock Brief Funct Genomic Proteomic 6: 302--312.
\bibAnnoteFile{koegl_improving_2007}

\bibitem{Gavin2006}
{Gavin} AC, {Aloy} P, {Grandi} P, {Krause} R, {Boesche} M, et~al. (2006)
  Proteome survey reveals modularity of the yeast cell machinery.
\newblock Nature 440: 631-636.
\bibAnnoteFile{Gavin2006}

\bibitem{Krogan2006}
{Krogan} NJ, {Cagney} G, {Yu} H, {Zhong} G, {Guo} X, et~al. (2006) Global
  landscape of protein complexes in the yeast saccharomyces cerevisiae.
\newblock Nature 440: 637-643.
\bibAnnoteFile{Krogan2006}

\bibitem{Wodak2009}
Wodak SJ, Pu S, Vlasblom J, S\'eraphin B (January 2009) Challenges and rewards
  of interaction proteomics.
\newblock Mol Cell Proteomics 8: 3-18.
\bibAnnoteFile{Wodak2009}

\bibitem{tarassov_vivo_2008}
Tarassov K, Messier V, Landry CR, Radinovic S, Serna~Molina MM, et~al. (2008)
  An in vivo map of the yeast protein interactome.
\newblock Science 320: 1465--1470.
\bibAnnoteFile{tarassov_vivo_2008}

\bibitem{Yu2008}
Yu H, Braun P, Y{\i}ld{\i}r{\i}m MA, Lemmens I, Venkatesan K, et~al. (2008)
  {High-quality binary protein interaction map of the yeast interactome
  network}.
\newblock Science 322: 104--110.
\bibAnnoteFile{Yu2008}

\bibitem{von_mering_comparative_2002}
von Mering C, Krause R, Snel B, Cornell M, Oliver SG, et~al. (2002) Comparative
  assessment of large-scale data sets of protein-protein interactions.
\newblock Nature 417: 399--403.
\bibAnnoteFile{von_mering_comparative_2002}

\bibitem{huang_where_2007}
Huang H, Jedynak BM, Bader JS (2007) Where have all the interactions gone?
  estimating the coverage of two-hybrid protein interaction maps.
\newblock PLoS Comput Biol 3: e214.
\bibAnnoteFile{huang_where_2007}

\bibitem{braun_experimentally_2009}
Braun P, Tasan M, Dreze M, Barrios-Rodiles M, Lemmens I, et~al. (2009) An
  experimentally derived confidence score for binary protein-protein
  interactions.
\newblock Nat Methods 6: 91--97.
\bibAnnoteFile{braun_experimentally_2009}

\bibitem{rajagopala_benchmarking_2009}
Rajagopala SV, Hughes KT, Uetz P (2009) Benchmarking yeast two-hybrid systems
  using the interactions of bacterial motility proteins.
\newblock Proteomics 9: 5296--5302.
\bibAnnoteFile{rajagopala_benchmarking_2009}

\bibitem{collins_toward_2007}
Collins SR, Kemmeren P, Zhao XC, Greenblatt JF, Spencer F, et~al. (2007) Toward
  a comprehensive atlas of the physical interactome of saccharomyces
  cerevisiae.
\newblock Mol Cell Proteomics 6: 439--450.
\bibAnnoteFile{collins_toward_2007}

\bibitem{rual_towards_2005}
Rual J, Venkatesan K, Hao T, {Hirozane-Kishikawa} T, Dricot A, et~al. (2005)
  Towards a proteome-scale map of the human protein-protein interaction
  network.
\newblock Nature 437: 1173--1178.
\bibAnnoteFile{rual_towards_2005}

\bibitem{ath_consortium_evidence_2011}
{Arabidopsis Interactome Mapping Consortium} (2011) {Evidence for network
  evolution in an arabidopsis interactome map}.
\newblock Science 333: 601--607.
\bibAnnoteFile{ath_consortium_evidence_2011}

\bibitem{licata_mint_2012}
Licata L, Briganti L, Peluso D, Perfetto L, Iannuccelli M, et~al. (2012)
  {MINT}, the molecular interaction database: 2012 update.
\newblock Nucleic Acids Res 40: D857--861.
\bibAnnoteFile{licata_mint_2012}

\bibitem{salwinski_database_2004}
Salwinski L, Miller CS, Smith AJ, Pettit FK, Bowie JU, et~al. (2004) The
  database of interacting proteins: 2004 update.
\newblock Nucleic Acids Res 32: D449--451.
\bibAnnoteFile{salwinski_database_2004}

\bibitem{breitkreutz_biogrid_2008}
Breitkreutz B, Stark C, Reguly T, Boucher L, Breitkreutz A, et~al. (2008) The
  {BioGRID} interaction database: 2008 update.
\newblock Nucleic Acids Res 36: D637--640.
\bibAnnoteFile{breitkreutz_biogrid_2008}

\bibitem{kerrien_intact_2012}
Kerrien S, Aranda B, Breuza L, Bridge A, {Broackes-Carter} F, et~al. (2012) The
  {IntAct} molecular interaction database in 2012.
\newblock Nucleic Acids Res 40: D841--846.
\bibAnnoteFile{kerrien_intact_2012}

\bibitem{szklarczyk_string_2011}
Szklarczyk D, Franceschini A, Kuhn M, Simonovic M, Roth A, et~al. (2011) The
  {STRING} database in 2011: functional interaction networks of proteins,
  globally integrated and scored.
\newblock Nucleic Acids Res 39: D561--568.
\bibAnnoteFile{szklarczyk_string_2011}

\bibitem{Barabasi2004}
Barabasi AL, Oltvai ZN (2004) {Network biology: understanding the cell's
  functional organization}.
\newblock Nat Rev Genet 5: 101--113.
\bibAnnoteFile{Barabasi2004}

\bibitem{Bollobas1985}
Bollob\'{a}s B (1985) Random graphs.
\newblock London: Academic Press.
\bibAnnoteFile{Bollobas1985}

\bibitem{Watts1998}
{Watts} DJ, {Strogatz} SH (1998) Collective dynamics of `small-world' networks.
\newblock Nature 393: 440-442.
\bibAnnoteFile{Watts1998}

\bibitem{Ravasz2002}
{Ravasz} E, {Somera} AL, {Mongru} DA, {Oltvai} ZN, {Barab{\'a}si} AL (2002)
  Hierarchical organization of modularity in metabolic networks.
\newblock Science 297: 1551-1555.
\bibAnnoteFile{Ravasz2002}

\bibitem{Song2005}
{Song} C, {Havlin} S, {Makse} HA (2005) Self-similarity of complex networks.
\newblock Nature 433: 392-395.
\bibAnnoteFile{Song2005}

\bibitem{albert_error_2000}
Albert, Jeong, Barabasi (2000) Error and attack tolerance of complex networks.
\newblock Nature 406: 378--382.
\bibAnnoteFile{albert_error_2000}

\bibitem{Albert2002}
Albert R, Barab\'asi AL (2002) Statistical mechanics of complex networks.
\newblock Rev Mod Phys 74: 47--97.
\bibAnnoteFile{Albert2002}

\bibitem{Barabasi1999}
Barab\'{a}si AL, Albert R (1999) {Emergence of scaling in random networks}.
\newblock Science 286: 509--512.
\bibAnnoteFile{Barabasi1999}

\bibitem{Wagner2001}
{Wagner} A (2001) The yeast protein interaction network evolves rapidly and
  contains few redundant duplicate genes.
\newblock Mol Biol Evol 18: 1283-1292.
\bibAnnoteFile{Wagner2001}

\bibitem{Song2006}
{Song} C, {Havlin} S, {Makse} HA (2006) Origins of fractality in the growth of
  complex networks.
\newblock Nat Physics 2: 275-281.
\bibAnnoteFile{Song2006}

\bibitem{Gallos2007}
{Gallos} LK, {Song} C, {Havlin} S, {Makse} HA (2007) Scaling theory of
  transport in complex biological networks.
\newblock Proc Natl Acad Sci 104: 7746-7751.
\bibAnnoteFile{Gallos2007}

\bibitem{Galvao2010}
Galvao V, Miranda JGV, Andrade RFS, Andrade JS, Gallos LK, et~al. (2010)
  Modularity map of the network of human cell differentiation.
\newblock Proc Natl Acad Sci 107: 5750-5755.
\bibAnnoteFile{Galvao2010}

\bibitem{Goh2006}
{Goh} KI, {Salvi} G, {Kahng} B, {Kim} D (2006) Skeleton and fractal scaling in
  complex networks.
\newblock Phys Rev Lett 96: 018701.
\bibAnnoteFile{Goh2006}

\bibitem{presser_evolutionary_2008}
Presser A, Elowitz MB, Kellis M, Kishony R (2008) The evolutionary dynamics of
  the saccharomyces cerevisiae protein interaction network after duplication.
\newblock Proc Natl Acad Sci 105: 950--954.
\bibAnnoteFile{presser_evolutionary_2008}

\bibitem{Ohno1970}
Ohno S (1970) {Evolution by gene duplication}.
\newblock Springer-Verlag.
\bibAnnoteFile{Ohno1970}

\bibitem{Li1997}
Li WS (1997) {Molecular evolution}.
\newblock Sunderland, MA: Sinauer Associates, Inc.
\bibAnnoteFile{Li1997}

\bibitem{Patthy1999}
Patthy L (1999) {Protein evolution}.
\newblock Portland, OR: Blackwell Publishers.
\bibAnnoteFile{Patthy1999}

\bibitem{Taylor2004}
Taylor JS, Raes J (2004) {Duplication and divergence: the evolution of new
  genes and old ideas}.
\newblock Annu Rev Genet 38: 615-643.
\bibAnnoteFile{Taylor2004}

\bibitem{wagner_how_2003}
Wagner A (2003) How the global structure of protein interaction networks
  evolves.
\newblock Proc Biol Sc 270: 457--466.
\bibAnnoteFile{wagner_how_2003}

\bibitem{qian_measuring_2011-1}
Qian W, He X, Chan E, Xu H, Zhang J (2011) Measuring the evolutionary rate of
  protein-protein interaction.
\newblock Proc Natl Acad Sci 108: 8725--8730.
\bibAnnoteFile{qian_measuring_2011-1}

\bibitem{Evlampiev2007}
Evlampiev K, Isambert H (2007) {Modeling protein network evolution under genome
  duplication and domain shuffling.}
\newblock BMC Syst Biol 1.
\bibAnnoteFile{Evlampiev2007}

\bibitem{Evlampiev2008}
{Evlampiev} K, {Isambert} H (2008) Conservation and topology of protein
  interaction networks under duplication-divergence evolution.
\newblock Proc Natl Acad Sci 105: 9863-9868.
\bibAnnoteFile{Evlampiev2008}

\bibitem{Sole2002}
{Sole} RV, {Pastor-Satorras} R, {Smith} E, {Kepler} TB (2002) A model of
  large-scale proteome evolution.
\newblock Adv Complex Syst 5: 43.
\bibAnnoteFile{Sole2002}

\bibitem{Kim2002}
{Kim} J, {Krapivsky} PL, {Kahng} B, {Redner} S (2002) Infinite-order
  percolation and giant fluctuations in a protein interaction network.
\newblock Phys Rev E 66: 055101.
\bibAnnoteFile{Kim2002}

\bibitem{Chung2002}
{Chung} F, {Lu} L, {Dewey} TG, {Galas} DJ (2002) Duplication models for
  biological networks.
\newblock J Comput Biol 10: 677.
\bibAnnoteFile{Chung2002}

\bibitem{Vazquez2003}
{Vazquez} A, {Flammini} A, {Maritan} A, {Vespignani} A (2003) Modeling of
  protein interaction networks.
\newblock Complexus 1: 38.
\bibAnnoteFile{Vazquez2003}

\bibitem{mirkin_algorithms_2003}
Mirkin BG, Fenner TI, Galperin MY, Koonin EV (2003) Algorithms for computing
  parsimonious evolutionary scenarios for genome evolution, the last universal
  common ancestor and dominance of horizontal gene transfer in the evolution of
  prokaryotes.
\newblock {BMC} evolutionary biology 3: 2.
\bibAnnoteFile{mirkin_algorithms_2003}

\bibitem{patro_parsimonious_2012}
Patro R, Sefer E, Malin J, Marçais G, Navlakha S, et~al. (2012) Parsimonious
  reconstruction of network evolution.
\newblock Algorithms for molecular biology: {AMB} 7: 25.
\bibAnnoteFile{patro_parsimonious_2012}

\bibitem{Navlakha2011}
Navlakha S, Kingsford C (2011) {Network archaeology: Uncovering ancient
  networks from present-day interactions}.
\newblock PLoS Comput Biol 7: e1001119.
\bibAnnoteFile{Navlakha2011}

\bibitem{zhang_refining_2010}
Zhang X, Moret BME (2010) Refining transcriptional regulatory networks using
  network evolutionary models and gene histories.
\newblock Algorithms for molecular biology: {AMB} 5: 1.
\bibAnnoteFile{zhang_refining_2010}

\bibitem{pinney_reconstruction_2007}
Pinney JW, Amoutzias GD, Rattray M, Robertson DL (2007) Reconstruction of
  ancestral protein interaction networks for the {bZIP} transcription factors.
\newblock Proceedings of the National Academy of Sciences of the United States
  of America 104: 20449--20453.
\bibAnnoteFile{pinney_reconstruction_2007}

\bibitem{gibson_reverse_2009}
Gibson TA, Goldberg DS (2009) Reverse engineering the evolution of protein
  interaction networks.
\newblock Pacific Symposium on Biocomputing Pacific Symposium on Biocomputing :
  190--202.
\bibAnnoteFile{gibson_reverse_2009}

\bibitem{Dutkowski2007}
Dutkowski J, Tiuryn J (2007) {Identification of functional modules from
  conserved ancestral protein protein interactions}.
\newblock Bioinformatics 23: i149--158.
\bibAnnoteFile{Dutkowski2007}

\bibitem{koonin_orthologs_2005}
Koonin EV (2005) Orthologs, paralogs, and evolutionary genomics.
\newblock Annu Rev Genet 39: 309--338.
\bibAnnoteFile{koonin_orthologs_2005}

\bibitem{Kunin2004}
Kunin V, Pereira-Leal JB, Ouzounis CA (2004) Functional evolution of the yeast
  protein interaction network.
\newblock Mol Biol Evo 21: 1171-1176.
\bibAnnoteFile{Kunin2004}

\bibitem{Muller2010}
Muller J, Szklarczyk D, Julien P, Letunic I, Roth A, et~al. (2010) {eggNOG
  v2.0: extending the evolutionary genealogy of genes with enhanced
  non-supervised orthologous groups, species and functional annotations.}
\newblock Nucleic Acids Res 38: D190--195.
\bibAnnoteFile{Muller2010}

\bibitem{Song2007}
{Song} C, {Gallos} LK, {Havlin} S, {Makse} HA (2007) How to calculate the
  fractal dimension of a complex network: the box covering algorithm.
\newblock J Stat Mech: Theory Exp 3: P03006.
\bibAnnoteFile{Song2007}

\bibitem{Rezenfeld2010}
{Rozenfeld} HD, {Song} C, {Makse} HA (2010) Small-world to fractal transition
  in complex networks: A renormalization group approach.
\newblock Phys Rev Lett 104: 025701.
\bibAnnoteFile{Rezenfeld2010}

\bibitem{Bunde1991}
Bunde A, Havlin S, editors (1996) Fractals and disordered systems.
\newblock New York: Springer-Verlag, 2st, edition.
\bibAnnoteFile{Bunde1991}

\bibitem{Jeong2001}
{Jeong} H, {Mason} SP, {Barab{\'a}si} AL, {Oltvai} ZN (2001) Lethality and
  centrality in protein networks.
\newblock Nature 411: 41-42.
\bibAnnoteFile{Jeong2001}

\bibitem{Amaral2000}
{Amaral} LAN, {Scala} A, {Barth{\'e}l{\'e}my} M, {Stanley} HE (2000) Classes of
  small-world networks.
\newblock Proc Natl Acad Sci 971: 11149-11152.
\bibAnnoteFile{Amaral2000}

\bibitem{jensen_eggnog_2008}
Jensen LJ, Julien P, Kuhn M, von Mering C, Muller J, et~al. (2008) {eggNOG:}
  automated construction and annotation of orthologous groups of genes.
\newblock Nucleic Acids Res 36: D250--254.
\bibAnnoteFile{jensen_eggnog_2008}

\bibitem{Gallos2012A}
Gallos LK, Makse HA, Sigman M (2012) A small world of weak ties provides
  optimal global integration of self-similar modules in functional brain
  networks.
\newblock Proc Natl Acad Sci 109: 2825-2830.
\bibAnnoteFile{Gallos2012A}

\bibitem{Granovetter1973}
Granovetter MS (1973) The strength of weak ties.
\newblock Am J Sociol 78: pp. 1360-1380.
\bibAnnoteFile{Granovetter1973}

\bibitem{gibson_questioning_2009}
Gibson TA, Goldberg DS (2009) Questioning the ubiquity of neofunctionalization.
\newblock PLoS Comput Biol 5: e1000252.
\bibAnnoteFile{gibson_questioning_2009}

\bibitem{Yang2008}
Yang L, Pei W, Li T, Cao Y, Shen Y, et~al. (2008) A fractal network model with
  tunable fractal dimension.
\newblock In: Neural Networks and Signal Processing, 2008 International
  Conference on. pp. 53 -57.
\bibAnnoteFile{Yang2008}

\bibitem{Bizhani2011}
{Bizhani} G, {Sood} V, {Paczuski} M, {Grassberger} P (2011) Random sequential
  renormalization of networks: Application to critical trees.
\newblock Phys Rev E 83: 036110.
\bibAnnoteFile{Bizhani2011}

\bibitem{sayers_database_2012}
Sayers EW, Barrett T, Benson DA, Bolton E, Bryant SH, et~al. (2012) Database
  resources of the national center for biotechnology information.
\newblock Nucleic Acids Res 40: D13--25.
\bibAnnoteFile{sayers_database_2012}

\bibitem{Gallos2012C}
Gallos LK, Barttfeld P, Havlin S, Sigman M, Makse HA (2012) Collective behavior
  in the spatial spreading of obesity.
\newblock Sci Rep 2: 454.
\bibAnnoteFile{Gallos2012C}

\bibitem{Girvan2002}
Girvan M, Newman MEJ (2002) Community structure in social and biological
  networks.
\newblock Proc Natl Acad Sci 99: 7821-7826.
\bibAnnoteFile{Girvan2002}

\bibitem{hedges_timetree_2006}
Hedges SB, Dudley J, Kumar S (2006) {TimeTree:} a public knowledge-base of
  divergence times among organisms.
\newblock Bioinformatics 22: 2971--2972.
\bibAnnoteFile{hedges_timetree_2006}

\end{thebibliography}


\clearpage

\section*{Figures}

\clearpage

\clearpage

\clearpage

\clearpage

\section*{Tables}

\begin{table}[!ht]
\caption{\bf{Organism overview.}}
\begin{tabular}{ |c| c| c| c| c| c| }
\hline
Organism name & Abbreviation & NCBI Taxonomy ID & $s_c^*$ & Nodes at $s_c^*$ & Interactions at $s_c^*$\\
\hline
Escherichia coli K-12 & eco & 83333 & 440 & 873 & 2321 \\
\hline
Saccharomyces cerevisiae & sce & 4932 & 980 & 2144 & 6000 \\
\hline
Arabidopsis thaliana & ath & 3702 & 400 & 727 & 905 \\
\hline
Caenorhabditis elegans & cel & 6239 & 560 & 485 & 438 \\
\hline
Drosophila melanogaster & dme & 7227 & 700 & 461 & 598 \\
\hline
Mus musculus & mmu & 10090 & 700 & 718 & 658 \\
\hline
Homo sapiens & hsa & 9606 & 700 & 1891 & 2840 \\
\hline
\end{tabular}
\begin{flushleft}
Overview of the organisms for which networks were reconstructed.
For each organisms the scientific name, three-letter-abreviaton
used in tables and figures, NCBI Taxonomy ID
\cite{sayers_database_2012}, filtering threshold $s_c^*$, node
count after filtering at $s_c^*$ and interaction count after
filtering at $s_c^*$ are shown.
\end{flushleft}
\label{tab:organisms}
\end{table}

\begin{table}[!ht]
\caption{\bf{Scaling exponents ($\gamma$, $d_B$, $d_M$) for the
different species.}}
\begin{tabular}{|c|c|c|c|c|c|c|}
\hline
Species  & $\gamma$ & $d_B$ & $d_M$ & Scale-free & Exponential & Fractal\\
\hline
eco & 1.9(1) & 3.6(3) & 1.3(4) & Yes & No & Yes\\
\hline
sce & $\infty$ & 3.0(2) & 1.5(1) & No & Yes & Yes\\
\hline
ath & $\infty$ & 1.5(1) & 2.1(2) & No & Yes & Yes\\
\hline
cel & 2.6(1) & 1.6(1) & 1.8(2) & Yes & No & Yes\\
\hline
dme & 3.0(1) & 1.6(1) & 1.3(2) & Yes & No & Yes \\
\hline
mmu  & 2.9(1) & 1.7(1) & 2.0(1)  & Yes & No & Yes\\
\hline
hsa & $\infty$ & 2.9(2) & 2.0(1) & No & Yes & Yes\\
\hline
\end{tabular}
\begin{flushleft}
According to the values of the scaling exponents, the seven species listed are grouped into two categories: scale-free fractal networks and exponential (non-scale-free) fractal
networks. The scale-free networks have a power-law degree distribution with exponent $\gamma$, and the non-scale-free fractal networks have an exponential degree distribution with
$\gamma \sim \infty$. Notice that none of the networks are small-world. Instead, they are characterized by fractal/modular structures.
\end{flushleft}
\label{tab:exponents_static}
\end{table}

\begin{sidewaystable}[!ht]
\caption{\bf{Fitting parameters in the duplication-divergence
model for all organisms.}}
\begin{tabular}{|c| c| c| c| c| c| c| c| c| c| c| c|}
\hline
Species & $p_x$ & $p_y$ &\multicolumn{9}{c|}{$\alpha(t)$} \\
\hline
&  & &prNOG & roNOG & maNOG & veNOG & inNOG & meNOG & fuNOG & KOG/euNOG & COG/NOG \\
\hline
eco & 0.7 & 0.0008 &  &  &  &  &  &  &  &  & 0.007 \\
\hline
sce & 0.7 & 0.0002 &  &  &  &  &  &  & 0.0008 & 0.0007 & 0.001 \\
\hline
ath & 0.7 & 0.0001 &  &  &  &  &  &  &  & 0.003 & 0.008 \\
\hline
cel & 0.5 & 0.0004 &  &  &  &  &  & 0.002 &  & 0.001 & 0.005 \\
\hline
dme & 0.5 & 0.0004 &  &  &  &  & 0.003 & 0.004 &  & 0.004 & 0.004 \\
\hline
mmu & 0.7 & 0.0002 &  & 0.001 & 0.001 & 0.001 &  & 0.001 &  & 0.001 & 0.003 \\
\hline
hsa & 0.7 & 0.0002 & 0.0002 &  & 0.0004 & 0.0005 &  & 0.0005 &  & 0.0003 & 0.0004 \\
\hline
\end{tabular}
\begin{flushleft}
$p_x$ and $p_y$ are time-independent and describe the probability
that an interaction is retained after a duplication and the
probability that an interaction is created \emph{de novo},
respectively. The fraction of interacting pairs in the ancestral
network at time $t$ is represented by $\alpha(t)$. There are in
total nine ancestral time levels for the organisms investigated:
the ancestral primates (prNOG), the ancestral rodents (roNOG),
the ancestral mammals (maNOG), the ancestral vertebrates (veNOG),
the ancestral insects (inNOG), the ancestral animals (meNOG), the
ancestral fungi (fuNOG), the ancestral eukaryotes (KOG/euNOG), and
the LUCA (COG/NOG). Existing time levels are specific for every species depending on its lineage.
\end{flushleft}
\label{tab:parameters}
\end{sidewaystable}
\clearpage

\begin{sidewaystable}[!ht]
\caption{\bf{Scaling exponents, growth rates and their relationships}}
\begin{tabular}{|c|c|c|c|c|c|c|c|c|c|c|c|}
\hline
 & \multicolumn{4}{c|}{static exponents} &  & \multicolumn{6}{c|}{dynamic growth rates} \\
\hline
Species & $d_B$ & $\gamma$ & $d_k$ & $1+d_B/d_k (=\gamma)$ &  & $r_N$ & $r_l$ & $r_k$ & $r_N/r_l (=d_B)$ & $1 + r_N/r_k (=\gamma)$ & $r_k/r_l (=d_k)$ \\
\hline
eco & 3.6(3) & 1.9(1) & 3.3(4) & 2.1(1) &  & 0.06 & 0.02 & 0.07 & 3 & 1.9 & 3.5 \\
\hline
sce & 3.0(2) & $\infty$ & 0.0(1) & $\infty$ &  & 0.23(3) & 0.07(1) & 0.0(1) & 3.3(8) & $\infty$ & 0 \\
\hline
mmus & 1.7(1) & 2.9(1) & 0.8(1) & 3.1(4) &  & 0.22(3) & 0.15(1) & 0.14(2) & 1.5(3) & 2.6(4) & 0.9(2) \\
\hline
hsa & 2.9(2) & $\infty$ & 0.0(2) & $\infty$ &  & 0.23(2) & 0.08(1) & 0.0(1) & 2.9(5) & $\infty$ & 0 \\
\hline
\end{tabular}
\begin{flushleft}
Scaling exponents ($d_B$, $\gamma$, $d_k$), growth rates ($r_N$, $r_l$, $r_k$) and their relationships derived from the dynamic analysis (The growth rates of \emph{E. coli} do not have
uncertainties because there are only two time levels). Here we selected the three largest networks (\emph{E. coli}, \emph{S. cerevisiae}, and \emph{H. sapiens}) and one sample
(\emph{M. musculus}) representing the smaller networks.
\end{flushleft}
\label{tab:exponents_dynamic}
\end{sidewaystable}
\clearpage

\appendix
\section*{Supporting Information}
\setcounter{figure}{0} \setcounter{table}{0}
 \makeatletter
\renewcommand{\thefigure}{S\@arabic\c@figure}
\renewcommand{\thetable}{S\@arabic\c@table} \makeatother

\begin{figure}[!ht]
\includegraphics[scale=0.5, angle=270]{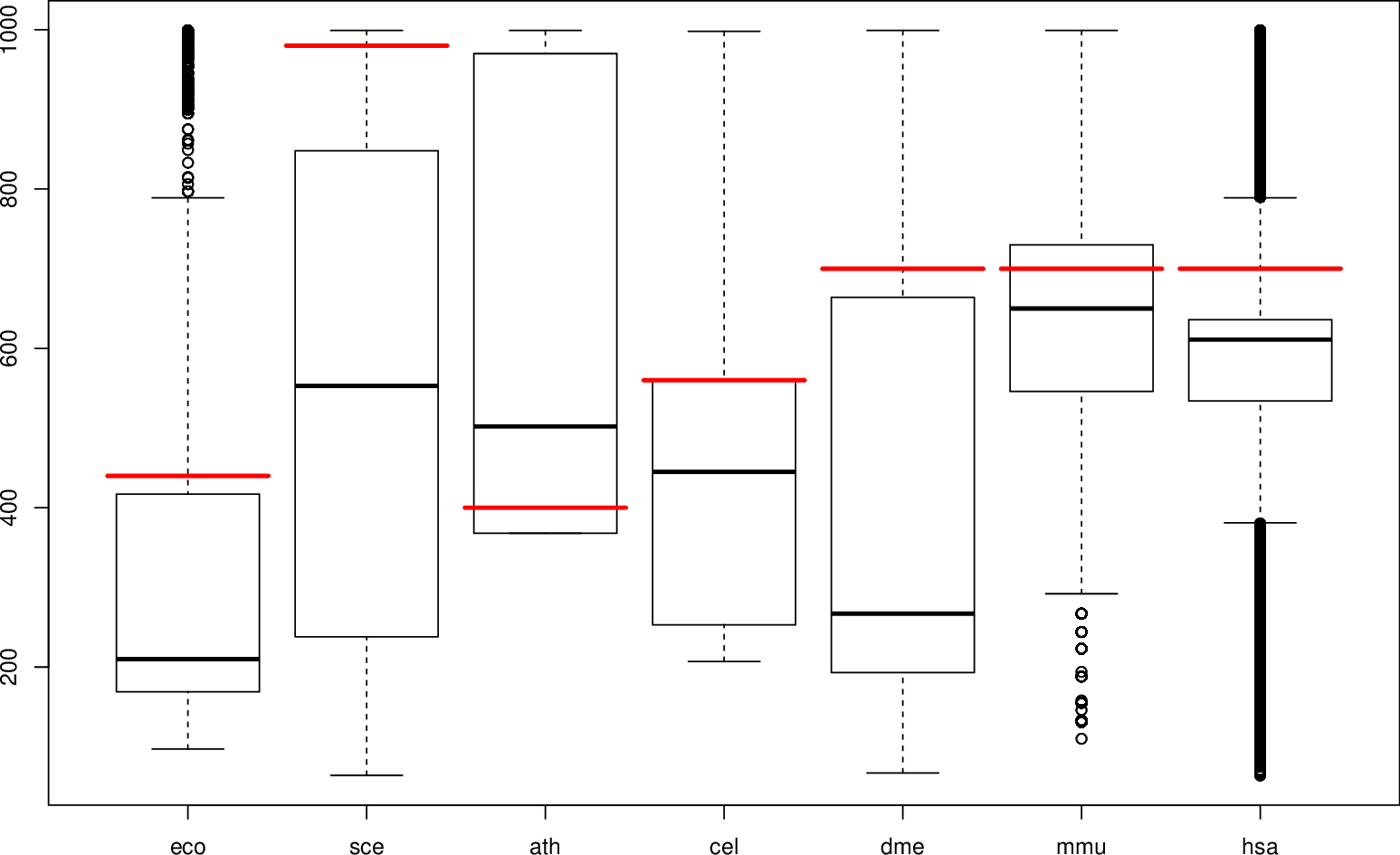}
\caption{{\bf Distribution of STRING experimental scores.} Box-and-whisker plots showing the distribution of STRING experimental scores for the organisms investigated. The filter
threshold $s_c^*$ for each species is indicated by a red line. The plots were created using the boxplot function of R.}
\label{fig:boxplot_allOrganisms_expScore}
\end{figure}

\begin{figure}[!ht]
\begin{center}
\hbox{A \includegraphics[width=4.5in]{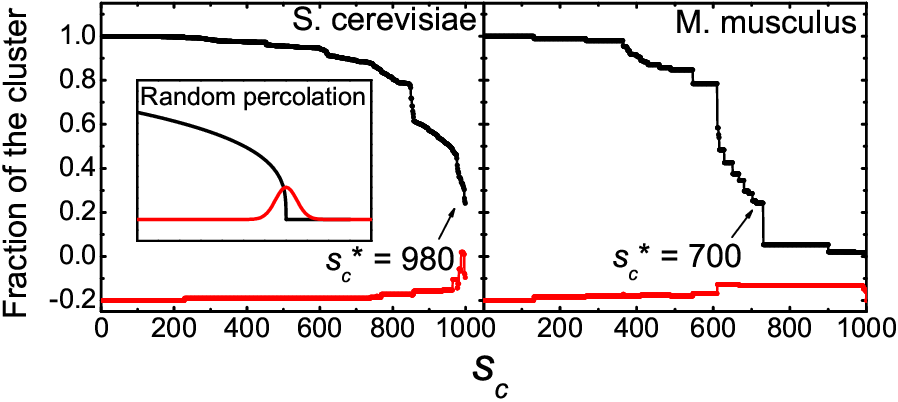}} \hbox{B
\includegraphics[width=4.5in]{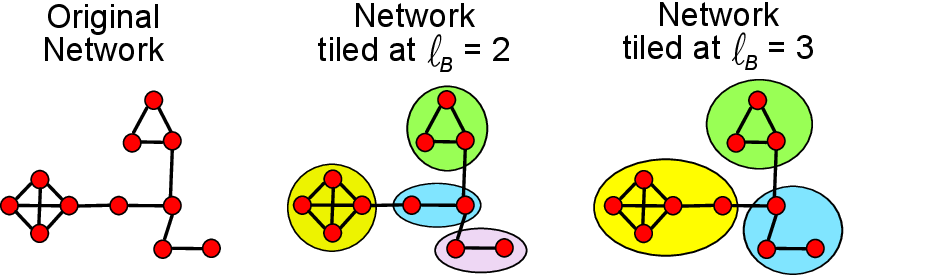}} \hbox{C
\includegraphics[width=4.5in]{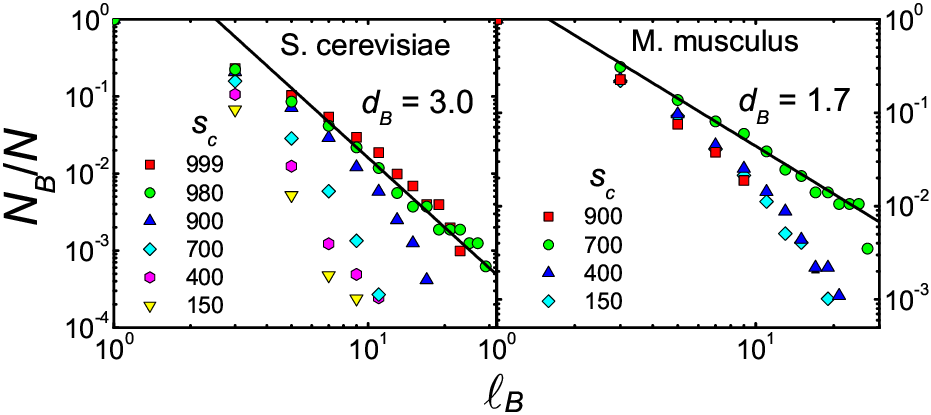}} \hbox{D
\includegraphics[width=4.5in]{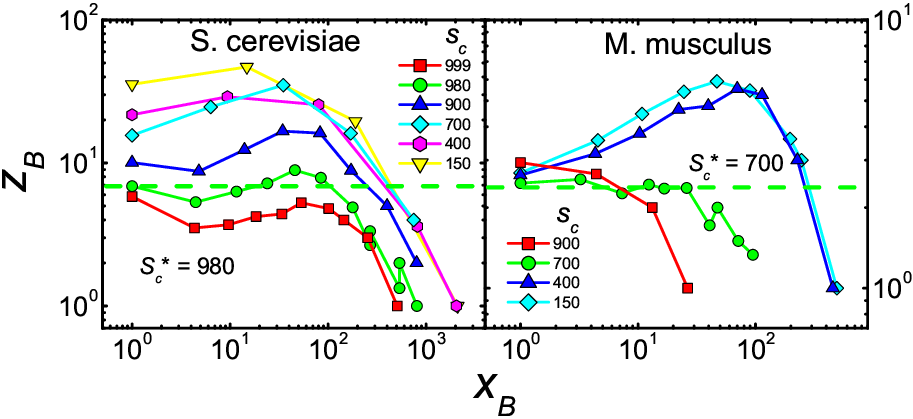}}
\end{center}
\caption{{\bf Determine the present-day PPI networks.} (A)
Percolation analysis of the present-day \textit{S. cerevisiae} and
\textit{M. musculus} PPI networks from the STRING database. We
plot the size of the largest (black) and second largest (red,
rescaled and shifted) connected components (as measured by the
fraction to the total number of nodes) versus cutoff score $s_c$.
The first jump of the largest connected component corresponds to
the threshold $s_c^*$. Inset shows schematically an uncorrelated
percolation. (B) Demonstration of the box-covering algorithm MEMB
\cite{Song2005, Song2007} for a schematic network. The network is
covered with boxes of size $\ell_B$. (C) Plot of the number of
boxes $N_B$ versus box size $\ell_B$ at different $s_c$. (D) $z_B$
versus $x_B$ under renormalization at different $s_c$. The dashed
line indicates the small-world to fractal transition point
$s_c^*$. } \label{fig:percolation}
\end{figure}
\begin{figure}[!ht]
\begin{center}
\hbox{\includegraphics[width=5in]{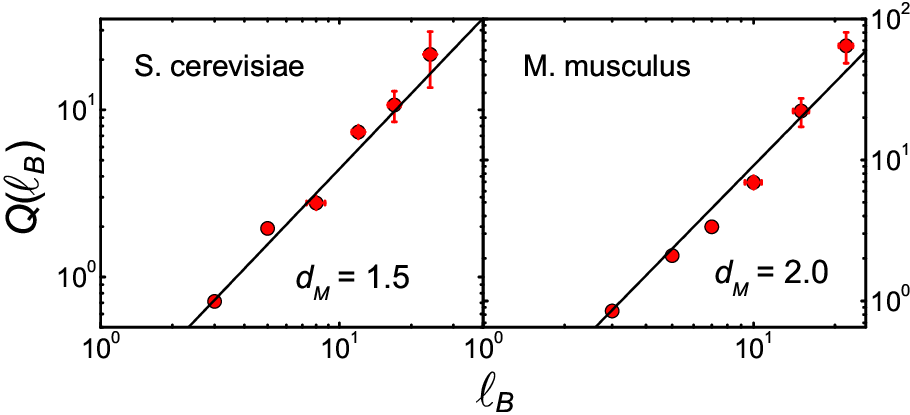}}
\end{center}
\caption{{\bf Modularity of PPI networks.} Log-log plot of the
modularity ratio $Q(\ell_B)$ versus size of the modules $\ell_B$.
Each point is an average over many modules with the same binned
$\ell_B$. The error bars are the standard deviations.}
\label{fig:modularity}
\end{figure}

\begin{figure}[!ht]
\hbox{A \includegraphics[scale=0.69, angle=-90]{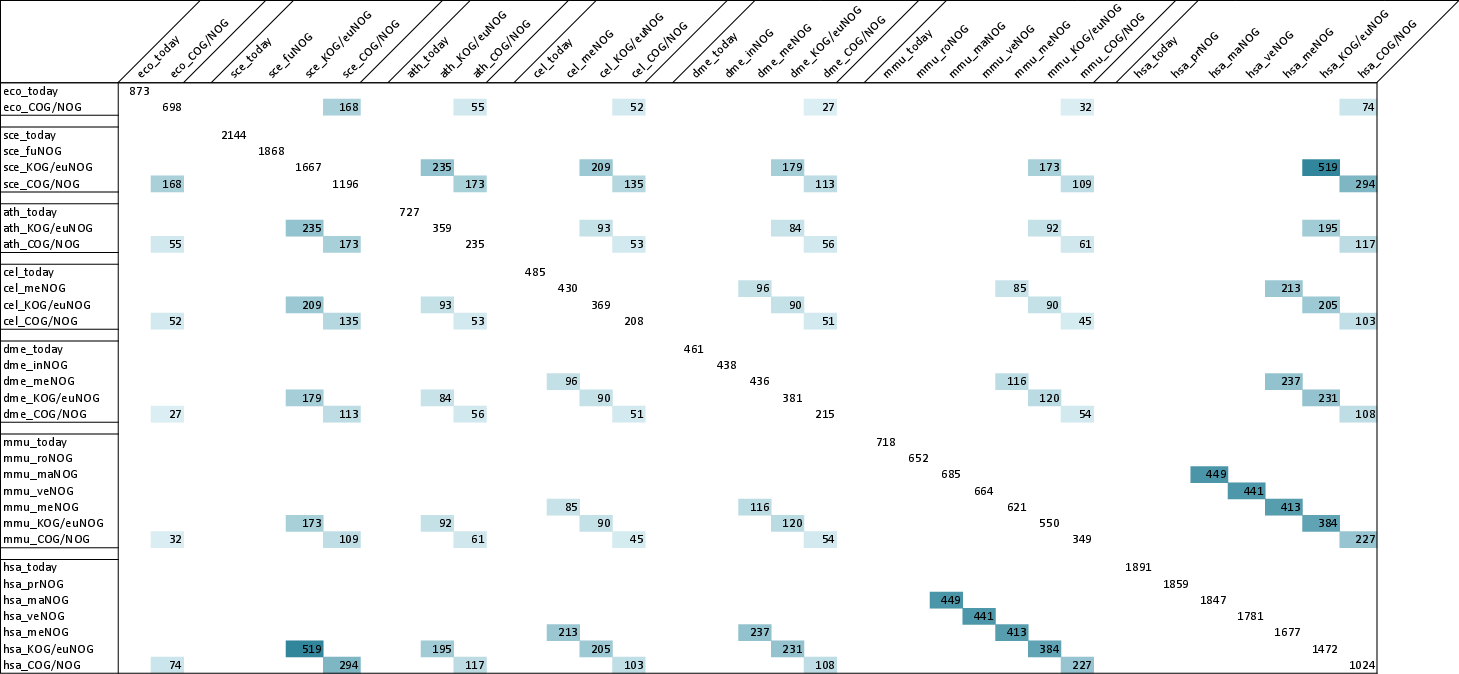}} \hbox{B
\includegraphics[scale=0.67,
angle=-90]{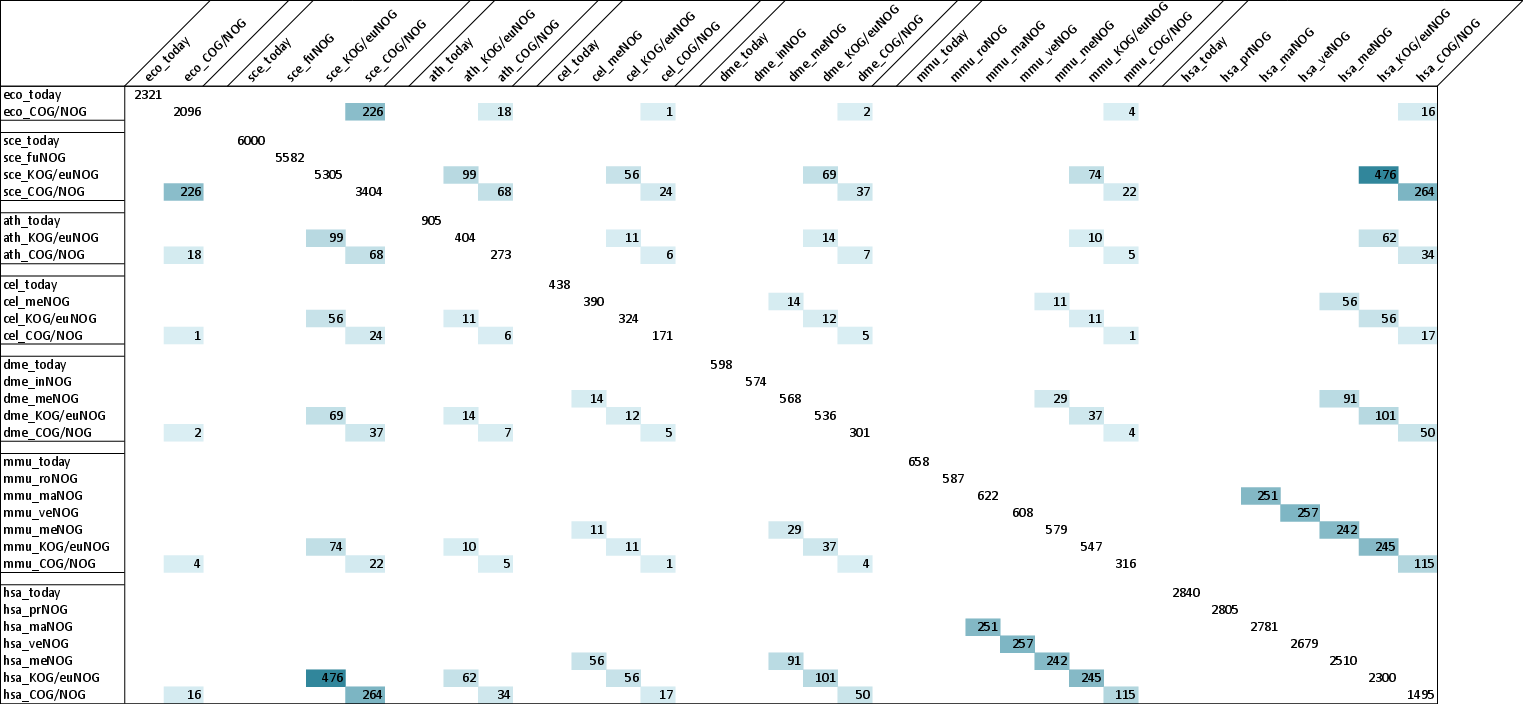}} \caption{{\bf Overlap of the different networks used for the study.} The overlaps between the networks of all organisms on all evolutionary
levels are shown, with the number of overlapping nodes in (A) and the number of overlapping interactions in (B). The color intensities represent the relative abundances in a heat
map-like manner, whith the lightest/darkest color referring to the lowest/highest number in the whole table except the diagonal. For example, while the interactome sizes are similar in
\textit{M. musculus} and \textit{A. thaliana}, the large overlap between the interactomes of \textit{H. sapiens} and \textit{M. musculus} can be attributed to their closer evolutionary
relationship. In case of equal evolutionary distances, the size of the interactome is decisive for the overlap; e.g. the overlap between \textit{E. coli} and \textit{S. cerevisiae} is
larger than the one between \textit{E. coli} and \textit{C. elegans}. In many cases, the overlaps in the ancient networks get smaller, which reflects the smaller network sizes.}

\label{fig:overlap_completeTable}
\end{figure}


\begin{figure}[!ht]
\begin{center}
\hbox{\includegraphics[width=5in]{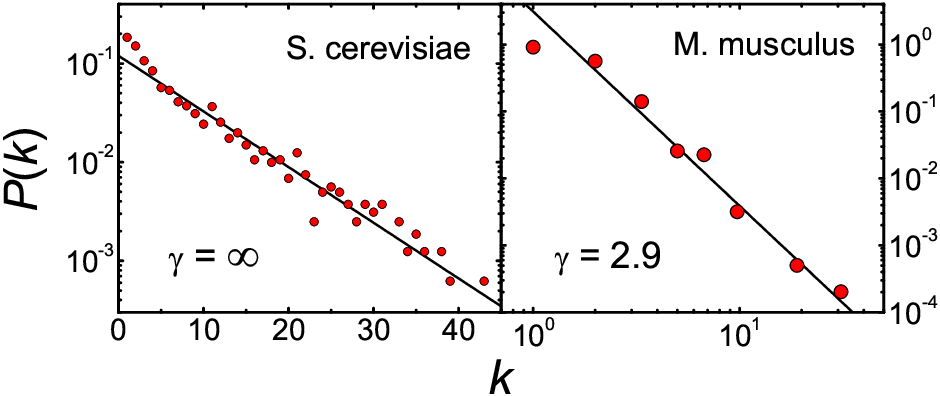}}
\end{center}
\caption{{\bf Degree distribution $P(k)$ of PPI networks.} Left, semi-log plot of $P(k)$ shows that the degree distribution of the \emph{S. cerevisiae} PPI network is exponential.
Right, log-log plot of $P(k)$ shows that the degree distribution of the \emph{M. musculus} PPI network is scale-free (power-law) with degree exponent $\gamma = 2.9(1)$.}
\label{fig:degree}
\end{figure}

\begin{figure}[!ht]
\hbox{ A
\includegraphics[width=2.5in]{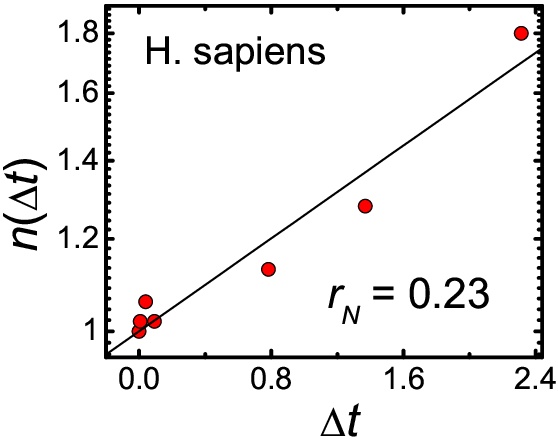}
       B \includegraphics[width=2.5in]{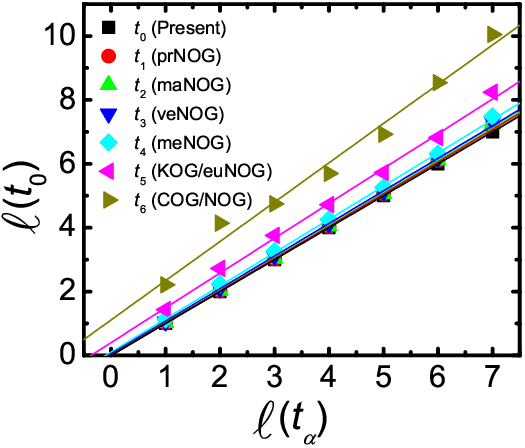}}
       \vspace{1cm}
\hbox{ C \includegraphics[width=2.5in]{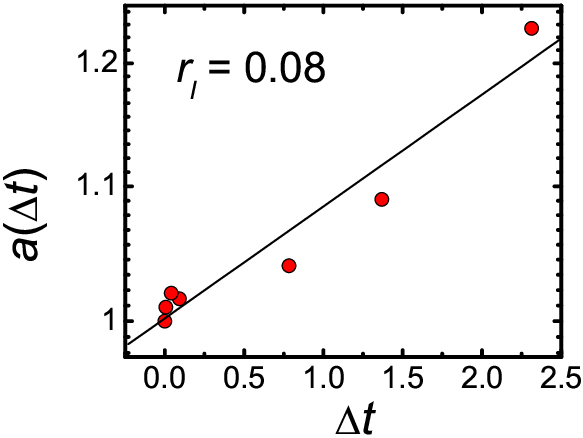}
       D \includegraphics[width=2.5in]{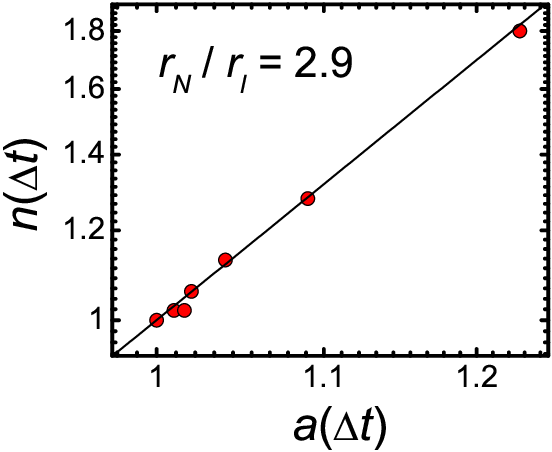}}
       \vspace{1cm}
\hbox{ E \includegraphics[width=2.5in]{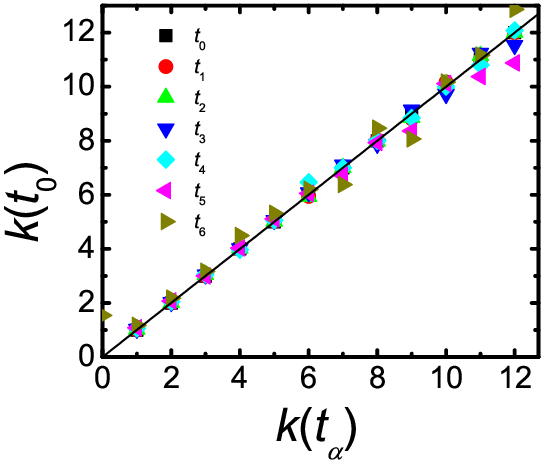}} \caption{{\bf Multiplicative growth mechanism of the \emph{H. sapiens} PPI network.} (A) $n(\Delta t)$ vs.
$\Delta t$. (B) Scaling between $\ell(t_0)$ and $\ell(t_\alpha)$. (C) $a(\Delta t)$ vs. $\Delta t$. (D) $n(\Delta t)$ vs. a($\Delta t$). (E) Scaling between $k(t_0)$ and $k(t_\alpha)$.
This figure is analogous to Figure~\ref{fig:dynamic_yeast} for \textit{S. cerevisiae}. } \label{fig:dynamic_human}
\end{figure}

\begin{figure}[!ht]
\hbox{ A
\includegraphics[width=2.5in]{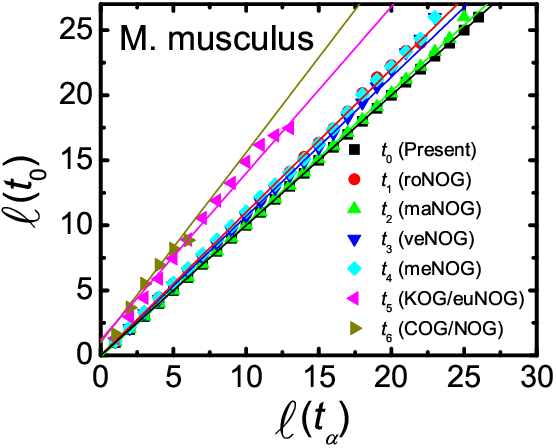}
       B \includegraphics[width=2.5in]{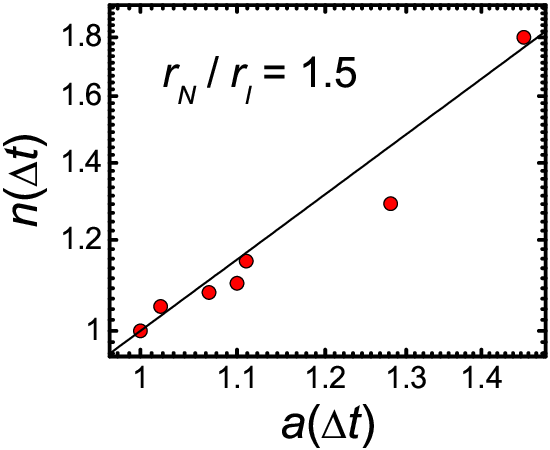}}
       \vspace{1cm}
\hbox{ C
\includegraphics[width=2.5in]{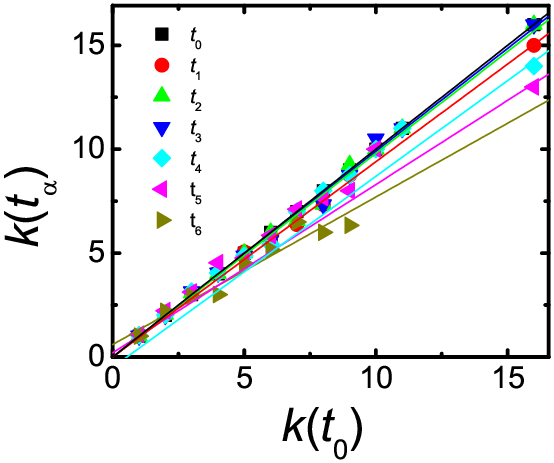}
       D \includegraphics[width=2.5in]{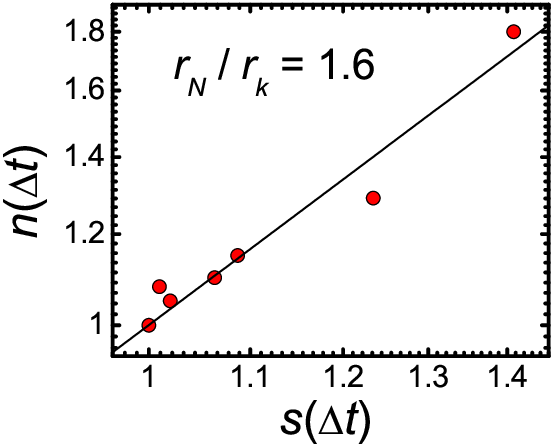}}
       \vspace{1cm}
\hbox{ E \includegraphics[width=2.5in]{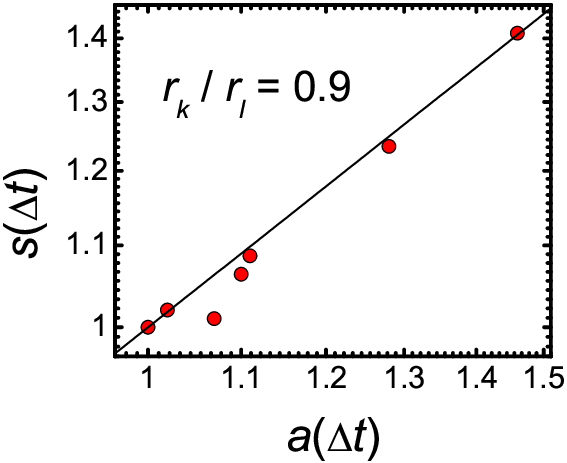}} \caption{{\bf Multiplicative growth mechanism of the \emph{M. musculus} PPI network.} (A) Scaling between $\ell(t_0)$ and
$\ell(t_\alpha)$. (B)  $n(\Delta t)$ vs. $a(\Delta t)$. (C)Scaling between $k(t_\alpha)$ and $k(t_0)$. (D) $n(\Delta t)$ vs. $s(\Delta t)$. (E) $s(\Delta t)$ vs. $a(\Delta t)$. This
figure is analogous to Figure~\ref{fig:dynamic_yeast} for \textit{S. cerevisiae}. Different from \textit{S. cerevisiae}, which has an exponential degree distribution, \textit{M.
musculus} has a power-law (scale-free) degree distribution (see Figure~\ref{fig:degree}).} \label{fig:dynamic_mouse}
\end{figure}

\begin{figure}[!ht]
\hbox{ A
\includegraphics[width=2.5in]{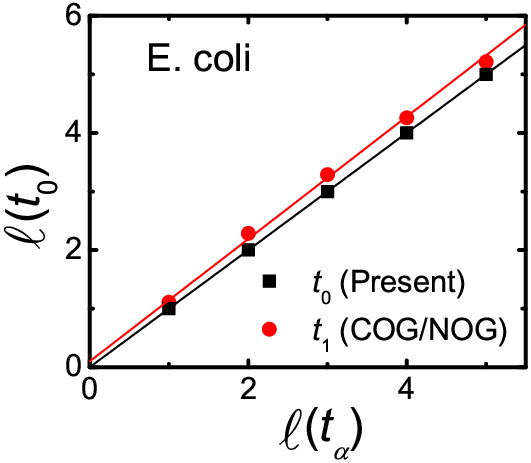}
       B \includegraphics[width=2.5in]{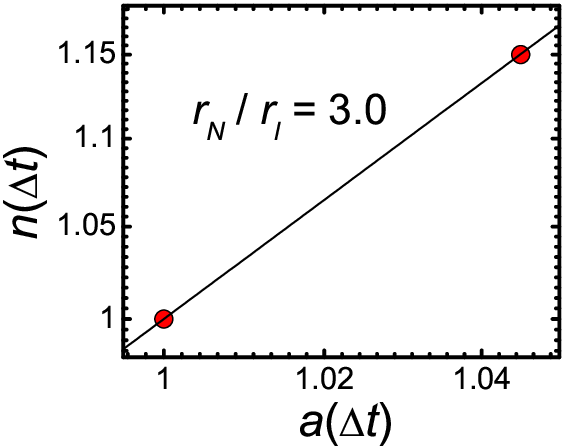}}
\vspace{1cm} \hbox{ C
\includegraphics[width=2.5in]{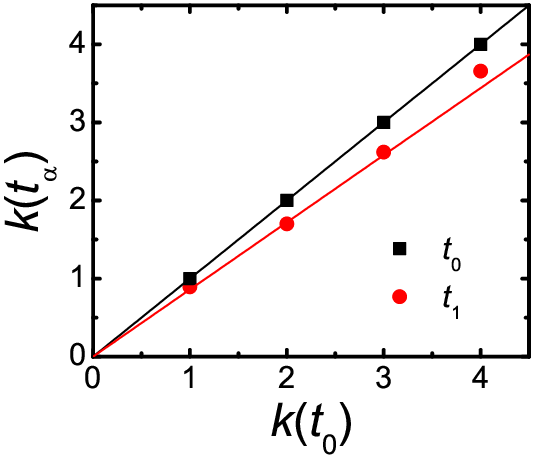}
       D \includegraphics[width=2.5in]{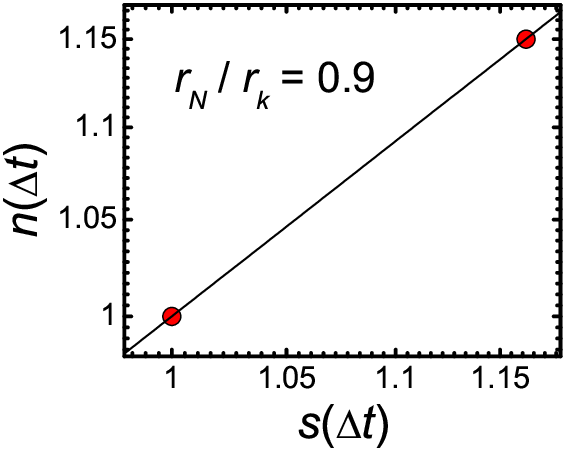}}
\vspace{1cm} \hbox{ E
\includegraphics[width=2.5in]{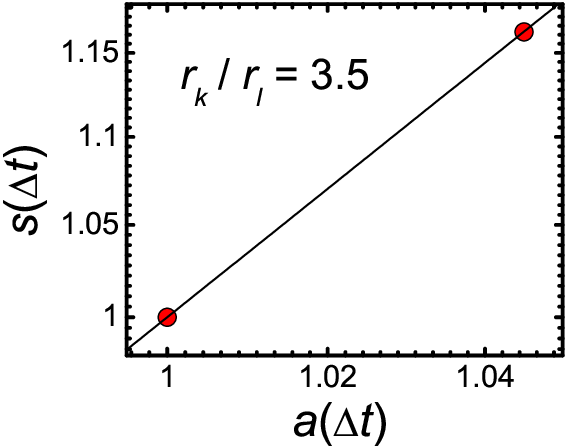}}
\caption{{\bf
Multiplicative growth mechanism of the \emph{E. coli} PPI
network.} (A) Scaling between $\ell(t_0)$ and $\ell(t_\alpha)$.
(B) $n(\Delta t)$ vs. $a(\Delta t)$. (C) Scaling between $k(t_0)$
and $k(t_\alpha)$. (D) $n(\Delta t)$ vs. $s(\Delta t)$. (E)
$s(\Delta t)$ vs. $a(\Delta t)$. This figure is analogous to
Figure~\ref{fig:dynamic_yeast} for \textit{S. cerevisiae}.}
\label{fig:dynamic_ecoli}
\end{figure}

\begin{figure}[!ht]
\begin{center}
\hbox{ A \includegraphics[width=2.5in]{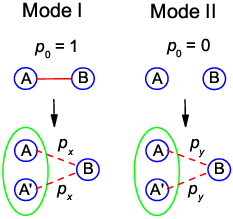} \hspace{1cm}  B
\includegraphics[width=2.5in]{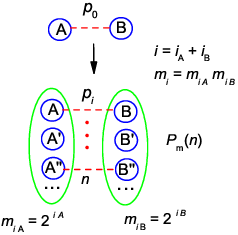}}
\vspace{1cm} C
\includegraphics[width=2.5in]{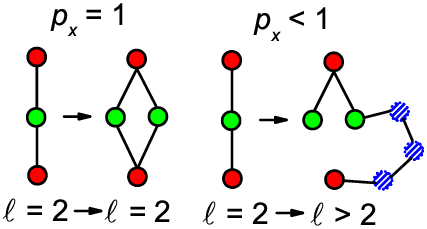}
\end{center}
\caption{{\bf Duplication-divergence model.} (A) The two basic
modes for the model. Left, mode I: protein A and B interact to
each other before duplication, and protein A duplicates to A and
A'. After duplication, A and A' have equal probability $p_x$ to
keep the interaction with B. Right, mode II: protein A and B do
not interact before duplication. After duplication, A and A' have
equal probability $p_y$ to generate a new interaction with B. (B)
Protein A and B duplicate to two clusters of $m_{i_A}$ and
$m_{i_B}$ proteins respectively after $i_A$ and $i_B$
duplications. We have $m_{i_A} = 2^{i_A}$, $m_{i_B} = 2^{i_B}$,
the total number of duplications $i = i_A + i_B$, and the total
number of possible links between cluster A and B $m_i = m_{i_A}
\times m_{i_B} = 2^i$. $p_n(m)$ is the probability to have $n$
interactions out of the $m$ total possible ones. (C) An example of
distance growth in the duplication-divergence model. Left,
distance $\ell$ between two proteins (red circles) does not change
when $p_x = 1$ (pure duplication of green circles, without
divergence). Right, $\ell$ increases when $p_x < 1$ due to the
loss of interactions. The red nodes are connected through a long
path of interactions between existing proteins (blue circles). }
\label{fig:model}
\end{figure}

\begin{figure}[!ht]
\begin{center}
\hbox{\includegraphics[width=5in]{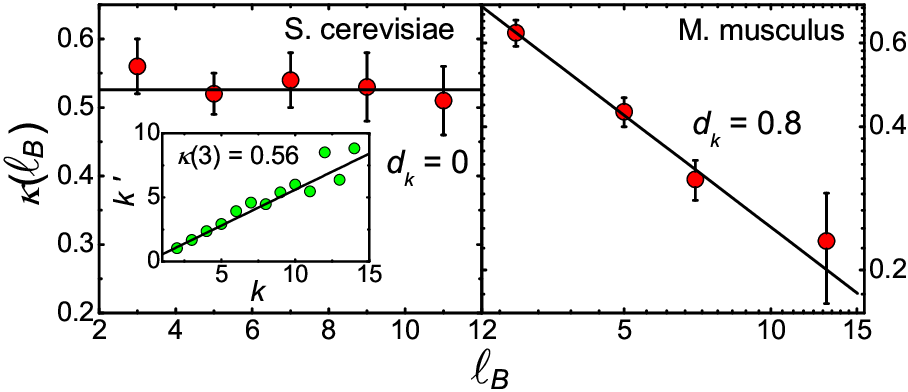}}
\end{center}
\caption{{\bf The scaling of $\kappa(\ell_B)$ vs. $\ell_B$.} The
renormalized degree exponent $d_k$ is calculated according to
Equation~(\ref{eq:degreeRG}). As an example, the inset shows the
renormalization relation $k' = \kappa(\ell_B)k$ for the case
$\ell_B = 3$. } \label{fig:dk}
\end{figure}

\begin{figure}[!ht]
\begin{center}
A \includegraphics[width=5in]{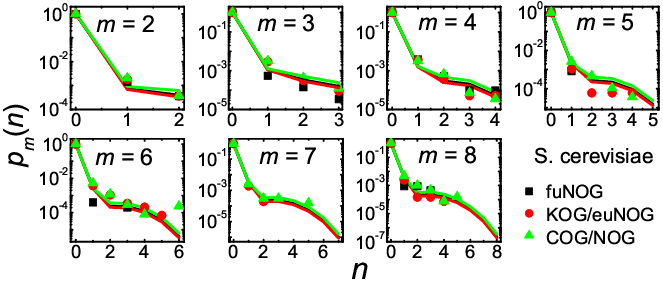} \hbox{B
\includegraphics[width=5in]{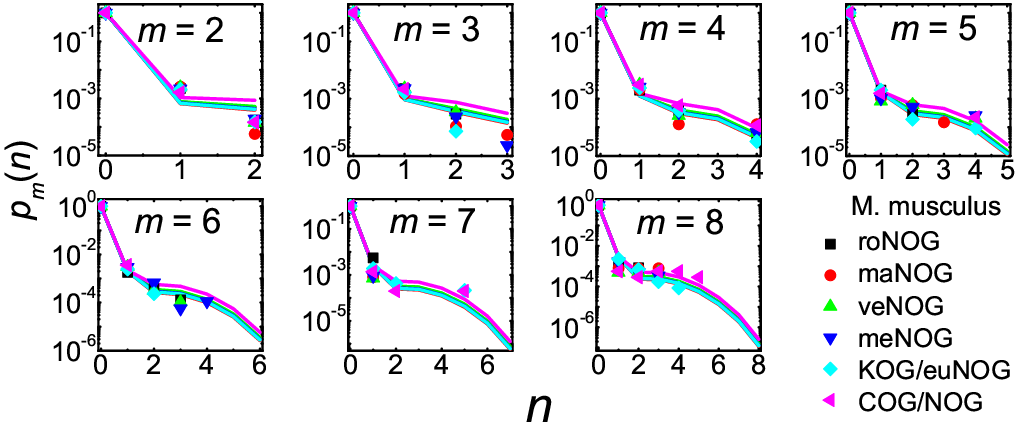}} \hbox{C
\includegraphics[width=5in]{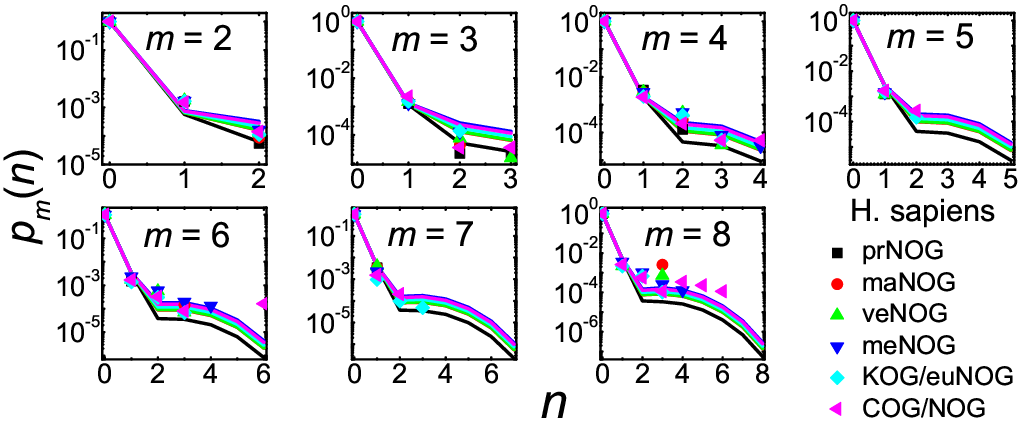}}
\end{center}
\caption{{\bf Fitting parameters and testing the duplication-divergence model.} Fit of $p_m(n)$ to the empirical data of (A) \textit{S. cerevisiae}, (B) \textit{M. musculus}, and (C) \textit{H. sapiens}. The curves are the fitted theoretical values, and the scatters are the empirical data. The model and the data are in good agreement. Parameters $p_x$, $p_y$ and $\alpha(t)$ (one $\alpha(t)$ for each time level $t$) of each species are determined from this fitting.} \label{fig:test}
\end{figure}

\begin{sidewaystable}[!ht]
\caption{{\bf Node and interaction counts at each filter step.} Numbers of proteins and interactions at each filter step preceding
the network construction and analysis. Four different filters were
applied: STRING experimental score $ > 0$, conservation on all
evolutionary levels defined for the corresponding organism in
eggNOG, filtering at the percolation threshold $s_c*$, and
filtering at the percolation threshold $s_c*$ and considering only
the largest connected component. The largest component (which is
also called giant component in the percolation literatures
\cite{Bunde1991}) is required for the topological analysis.}

\begin{tabular}{|c| c| c| c| c| c| c| c| c|}
\hline
   & \multicolumn{2}{c|}{Exp score $ > 0$ } & \multicolumn{2}{c|}{Conserved on all eggNOG levels} & \multicolumn{2}{c|}{After filtering at $s_c*$} & \multicolumn{2}{c|}{After filtering at $s_c*$ largest component} \\
\hline
Species & proteins & interactions & proteins & interactions & proteins & interactions & proteins & interactions \\
\hline
eco & 2472 & 11016 & 2472 & 11016 & 873 & 2321 & 705 & 2209 \\
\hline
sce & 5388 & 124956 & 4197 & 75625 & 2144 & 6000 & 1609 & 5546 \\
\hline
ath & 1913 & 3513 & 1104 & 1792 & 727 & 905 & 404 & 618 \\
\hline
cel & 3370 & 6768 & 997 & 1391 & 485 & 438 & 249 & 271 \\
\hline
dme & 5376 & 8695 & 2213 & 2915 & 461 & 598 & 311 & 504 \\
\hline
mmu & 3513 & 5623 & 1321 & 1573 & 718 & 658 & 285 & 351 \\
\hline
hsa & 10617 & 51573 & 5445 & 19040 & 1891 & 2840 & 1365 & 2435 \\
\hline
\end{tabular}
\label{tab:completeInputdataOverview}
\end{sidewaystable}

\begin{table}[!ht]
\begin{center}
\caption{{\bf Divergence times.} Estimated divergence times for the evolutionary levels in the eggNOG database. They represent the time point when the last common ancestor of a certain
evolutionary level existed. Estimates are derived from the TimeTree database \cite{hedges_timetree_2006}.}
\begin{tabular}{ |c| c|}
\hline
eggNOG level &  Divergence time (million years)\\
\hline
COG/NOG & 2313.2 \\
\hline
KOG/euNOG & 1369 \\
\hline
fuNOG & 798 \\
\hline
meNOG & 782.7 \\
\hline
inNOG & 366 \\
\hline
veNOG & 400.1 \\
\hline
maNOG & 92.4 \\
\hline
roNOG & 25.2 \\
\hline
prNOG & 6.4 \\
\hline
\end{tabular}
\label{tab:divergenceTimes}
\end{center}
\end{table}

\end{document}